\documentclass[aps,prd,superscriptaddress,nofootinbib,amsmath,amsfonts,preprintnumbers,groupedaddress,
showpacs,10pt,english]{revtex4}
\usepackage{amsmath}
\usepackage{amssymb}
\usepackage{babel}
\usepackage{wrapfig}
\usepackage{cancel}

\makeatletter

\usepackage{array,multirow,graphicx}
\usepackage{dcolumn}
\usepackage{newlfont}
\usepackage{bm}
\usepackage[colorlinks,citecolor=blue,urlcolor=blue,linkcolor=blue]{hyperref}
\usepackage[figtopcap]{subfigure}
\usepackage{color}

\begin{document}

\date{\today}

\title{General form of \(f(R)\) and $(2+1)$-dimensional charge/non-charge black hole solutions}

\author{G.G.L. Nashed}
\email{nashed@bue.edu.eg}
\affiliation {Centre for Theoretical Physics, The British University in Egypt, P.O. Box
43, El Sherouk City, Cairo 11837, Egypt}\affiliation {Centre for Space Research, North-West University, Potchefstroom 2520, South Africa}

\begin{abstract}
We introduce novel black hole (BH) solutions, charge/non-charge, within the framework of $f(R)$ gravity, a theory that does not inherently include a cosmological constant, using equal/diffirent metric ansatzs. Remarkably, these solutions exhibit asymptotically Anti-de Sitter (AdS) or de Sitter (dS) behavior, depending on their parameter values. Unlike the BTZ solutions of General Relativity, which feature a causal singularity and constant scalar invariants, our solutions display strong spacetime singularities, as shown by their scalar invariants. We construct $f(R)$ functions that behave as polynomial functions, emphasizing the unique nature of these solutions. We demonstrate the stability of these solutions in two ways: first, by showing that their heat capacity is positive, which ensures thermodynamic stability; and second, by proving that the second derivative of $f(R)$ is positive, meeting the Ostrogradski criterion for dynamical stability. Furthermore, the solutions satisfy the first law of thermodynamics, confirming their consistency with fundamental thermodynamic principles.
\end{abstract}


\maketitle
%%%%%%%%%%%%%%%%%%%%%%%%%%%%%%%%%%%%%%%%%%%%%%%%%%%%%%%%%%%%%%%%%%%%%%%%%%%%%%%%%%%%%%%%%%
\section{\bf Introduction}
Gravity continues to pose profound mysteries, yet Einstein’s General Relativity (GR) has provided an exceptionally successful description of gravitational phenomena. The theory has been subjected to rigorous testing, spanning from millimeter-scale laboratory experiments to observations across the solar system \cite{Will:2014kxa}. All such tests have consistently upheld the predictions of GR, including the observed gravitational radiation emitted by binary pulsars.

A pivotal confirmation of GR was realized in 2017 with the detection by Advanced LIGO and Virgo of GW170817, the first observed binary neutron star merger \cite{LIGOScientific:2017vwq}. As expected from GR these systems emit polarized gravitational waves, thereby offering further support for the theory.

The standard cosmological framework relies on GR as the core theory governing gravity across all scales. In 1998, observational astronomy underwent a paradigm shift with the discovery that the Universe is not decelerating, but rather accelerating in its expansion \cite{riess1998observational,SupernovaCosmologyProject:1998vns}. This conclusion was strongly supported by type Ia supernova surveys \cite{SupernovaSearchTeam:1998bnz,HST:2000azd,SNLS:2005qlf,SupernovaCosmologyProject:2008ojh}, detailed measurements of the Cosmic Microwave Background \cite{Boomerang:2000efg,Balbi:2000tg,Pryke:2001yz,WMAP:2003elm,Hou:2012xq,ade2016planck,Story:2012wx,AtacamaCosmologyTelescope:2013swu}, and studies of large-scale cosmic structures \cite{SDSS:2003eyi,SDSS:2004kqt,SDSS:2005xqv,Blake:2011en,Beutler:2011hx,BOSS:2012dmf}. Altogether, these findings provide strong support for the $\Lambda$CDM model \cite{younesizadeh2020modified,Bahcall:1999xn}.

In recent decades, the accelerated expansion of the Universe has stood as one of the most pressing problems in theoretical physics. Within GR, this effect is attributed to dark energy, a mysterious form of energy permeating space \cite{clifton2012modified}. The simplest representation of dark energy is the cosmological constant, which Einstein originally added to his equations when they implied a dynamic rather than static universe. Yet, the value required to explain the present acceleration is extremely small. From particle physics, vacuum energy is expected to generate a cosmological constant, but its theoretical estimate exceeds the observed value by more than 50 orders of magnitude. This enormous discrepancy highlights the possibility that the explanation for accelerated expansion may lie in modifications of gravity at cosmic scales..

  In the last few decades, research into modified gravity has flourished, positioning such models as credible alternatives to the concept of dark energy. These theories introduce new ideas to tackle the cosmological constant problem and explain the Universe’s accelerated expansion. Theoretical approaches to dark energy are typically grouped into two categories \cite{copeland2006dynamics,sahni20045,Sahni:2004ai}. The first introduces exotic matter components, such as scalar fields—quintessence and k-essence—that could account for acceleration. The second involves modifying the laws of gravity themselves, leading to frameworks like $f(R)$ gravity, scalar–tensor theories, and brane-world models. Each provides a unique way of rethinking gravity on cosmological scales, potentially resolving the mysteries surrounding dark energy and cosmic acceleration..

 The prediction of black holes represents a major milestone in GR, a theory that brings forth intriguing and unconventional ideas in physics. The thermodynamic nature of black holes became evident through the pioneering contributions of Hawking and Bekenstein \cite{Hawking:1974rv,Bekenstein:1973ur,Hawking:1975vcx}. This ground-breaking study revealed that black holes possess thermodynamic properties, including entropy and temperature, which are inherently connected to their horizon area and surface gravity both of which are fundamentally geometric in nature \cite{Bardeen:1973gs}. Moreover, black hole thermodynamics involves various critical phenomena, deepening our insight into these mysterious entities. One extensively studied topic in this context is the Hawking-Page (HP) phase transition \cite{Carlip:1993sa,Mann:1996ze,Maloney:2007ud}. This phenomenon is closely tied to global stability, which depends on the sign of the Gibbs free energy. A black hole is deemed globally stable when its Gibbs free energy remains positive at positive temperatures. The HP phase transition takes place at the exact points where the Gibbs free energy reaches zero.  Beyond global stability, another key concept is local stability, which has attracted considerable attention in recent years. This form of stability is directly linked to a black hole's heat capacity, where a positive value signifies thermodynamic stability, while a negative value indicates instability. In this framework, two types of phase transitions are recognized: Type one transitions occur at points where the heat capacity vanishes, whereas type two transitions take place when the heat capacity diverges \cite{Myung:2007my,Carter:2005uw,Capela:2012uk}.

The study of black hole thermodynamics reveals complex phase structures and critical behavior that mirror those of ordinary thermodynamic systems, most notably the van der Waals liquid–gas model \cite{Shen:2005nu,Kubiznak:2012wp}. In this analogy, the cosmological constant acts as thermodynamic pressure, with its conjugate variable representing thermodynamic volume \cite{Gibbons:1996af}. In Anti–de Sitter spacetime, black holes undergo phase transitions whose critical phenomena resemble those of the van der Waals fluid \cite{Dolan:2010ha}. Progress in this field was significantly advanced by the discovery of the BTZ black hole in $(2+1)$ dimensions by Bañados, Teitelboim, and Zanelli (1992) \cite{Banados:1992wn}. Gravity in three dimensions has particular appeal due to its relevance for quantum gravity, string theory, gauge theories, and the AdS/CFT and AdS/CMT correspondences \cite{Witten:1998qj,Nashed:2011fz,witten1998anti,witten1998anti}, while also offering technical simplifications in solving Einstein’s equations. In this paper, we highlight the differences between the BTZ black hole and our proposed $(2+1)$-dimensional solutions, both in structure and thermodynamic behavior. Constructing charged black hole solutions in GR is notoriously challenging because of the nonlinear character of Einstein’s field equations. The first exact solution of this type was the Reissner–Nordström metric \cite{Reissner:1916cle,1918KNAB...20.1238N}, describing the external field of a charged source in four dimensions. Building on this tradition, we explore new $(2+1)$-dimensional solutions in $f(R)$ gravity and present a systematic discussion of their properties in the following sections.

The structure of the present study is  as follows: Section~\ref{S2} introduces the fundamental aspects of $f(R)$ gravity and formulates its corresponding  equations of motion  in the presence of an electromagnetic field. In Section~\ref{S3}, we investigate these field equations within a (2+1)-dimensional spacetime characterized by two undetermined metric functions, $k(r)$ and $k_1(r)$. The equations are categorized into six distinct scenarios: ({\bf i}) constant $f_R$, with $k_1=1$ and zero electric charge; ({\bf ii}) constant $f_R$, $k_1=1$, and non-zero charge; ({\bf iii}) variable $f_R$, $k_1=1$, with vanishing charge; ({\bf iv}) variable $f_R$, $k_1=1$, and a non-zero charge; ({\bf v}) variable $f_R$, $k_1 \neq 1$, and zero charge; and ({\bf vi}) variable $f_R$, $k_1 \neq 1$, with non-zero charge.\footnote{Here, $f_R \equiv \frac{df(R)}{dR}$, and we consider its dependence on the radial coordinate r due to the assumption of spherical symmetry. The chain rule $f_R = \frac{df(R)}{dr} \cdot \frac{dr}{dR}$ is employed accordingly.}

It is important to note that the first two cases reproduce known solutions within Einstein's GR, including the BTZ black hole with and without electric charge, and therefore offer no novel results. However, the remaining four cases lead to new analytical solutions, which we explore in detail along with their asymptotic behavior. A striking result is that although our model does not assume a cosmological constant, the derived solutions naturally exhibit asymptotically (anti)-de Sitter geometry.

We investigate the role of $f(R)$ corrections by evaluating the main curvature invariants, namely the Ricci scalar, the Ricci tensor squared, and the Kretschmann scalar. The results reveal that the inclusion of these modifications strengthens the singularity in comparison with the standard $(2+1)$-dimensional black hole solutions in GR. In Section~\ref{BH}, explicit expressions for $f(R)$ and its first- and second-order derivatives are obtained for the last four cases, and their properties are examined through graphical analysis. The study shows that these functions remain positive, which indicates the absence of Ostrogradski-type instabilities. Section~\ref{S5} is devoted to the thermodynamic aspects of cases ({\bf iii})–({\bf vi}), where entropy, Hawking temperature, and heat capacity are computed. Final section is devoted for conclusion and  outlines the main results of this work.

%%%%%%%%%%%%%%%%%%%%%%%%%%%%%%%%%%%%%%%%%%%%%%%%%%%%%%%%%%%%%%%%%%%%%%%%%%%%%%%%%%%%%%
\section{Essentials of $f(R)$ Theory}\label{S2}
In this section, we focus on the $(2+1)$-dimensional framework of $f(R)$  and formulate its   equations of motion. It should be stressed that $f(R)$ gravity is a generalization of Einstein’s theory, which is retrieved when the function is chosen as $f(R)=R$. For any case where $f(R)\neq R$, the dynamics deviate from standard GR. The action describing the $f(R)$ model is given by \cite{Carroll:2003wy,1970MNRAS.150....1B,Nojiri:2003ft,Capozziello:2003gx,Capozziello:2011et,Nojiri:2010wj,Nojiri:2017ncd,Capozziello:2002rd}:
\begin{eqnarray} \label{a2}
I_G=\frac{1}{2\kappa^2} \int \sqrt{-g}  f(R)d^3x  +\int \sqrt{-g} {\cal L}_{ em}~d^{3}x. \  \
\end{eqnarray}
In this formulation, we take $\kappa^{2}=8\pi G$, with $G$ being Newton’s  constant, while $g$ denotes the determinant of the metric tensor. Electromagnetic sector in Eq.~(\ref{a2}) is described by the Lagrangian 
${\cal L}_{em}=-\tfrac{1}{2}\,F \wedge ^{\star}F$. The field strength is defined as $F=dA$, where the 1-form potential is written as $A=A_{\mu}dx^{\mu}$~\cite{Nashed:2024ush,Nashed:2005kn,Capozziello:2023vvr}\footnote{Electric charge is an inherent attribute of matter that determines its electromagnetic interactions. In classical physics, it is regarded as a conserved quantity in isolated systems, a property that follows from the symmetry principles underlying physical laws~\cite{Landau:1975pou}. Noether’s theorem establishes that every continuous symmetry of the action corresponds to a conserved quantity; charge conservation, in particular, stems from $U(1)$ gauge invariance within quantum field theory. Hidden or additional symmetries could, in principle, imply further conservation laws, potentially linked to phenomena not yet fully understood~\cite{Peskin:1995ev}. In more advanced approaches, such as Grand Unified Theories (GUTs) or extensions of the Standard Model, the concept of charge conservation is often interpreted as a manifestation of deeper and more fundamental symmetries.}.

The field equations are obtained by varying Eq.~(\ref{a2}) separately with respect to the metric tensor and the electromagnetic gauge 1-form. This procedure gives \cite{2005JCAP...02..010C}:
\begin{align} \label{f1}
\mathit{ R}_{\mu \nu} \mathit{f_{R}}-\frac{1}{2}g_{\mu
\nu}\mathit{f( R)}+[g_{\mu \nu}\nabla^2 -\nabla_\mu
\nabla_\nu]\mathit{ f}_{_\mathit{ R}}&=-\frac{\kappa^2}{2}{{{\cal
T}^{{}^{{}^{^{}{\!\!\!\!\scriptstyle{em}}}}}}}_{\mu \nu}\,,\end{align}
\begin{align}\label{e1}
\partial_\nu \left( \sqrt{-g} F^{\mu \nu} \right)&=0\;,\end{align}
with $\nabla^2 \equiv \nabla_\mu \nabla^\mu$.  

If we now contract Eq.~(\ref{f1}) in $(2+1)$ dimensions, the trace relation follows as
\begin{eqnarray} \label{f3}
2\nabla^2 {\mathit f_{R}}+\mathit{ R}{f_{R}}-\frac{3 \mathit
f(R)}{2}=0 \,.
\end{eqnarray}

From this trace equation, one can reformulate the function $\mathit f(R)$ in $(2+1)$ dimensions as
\begin{eqnarray} \label{f3s}
\mathit f(R)=\frac{2}{3}\Big[2\nabla^2 {\mathit f_{R}}+\mathit{
R}{f_{R}}\Big]\,.\end{eqnarray}

Inserting Eq.~(\ref{f3s}) into Eq.~(\ref{f1}) yields the modified field equations:
\begin{eqnarray} \label{f3ss}
\mathit{ R}_{\mu \nu} \mathit{f_{R}}-\frac{1}{3}g_{\mu
\nu}\mathit{ R}\mathit{ f}_{_\mathit{ R}}+\frac{1}{3}g_{\mu
\nu}\nabla^2\mathit{ f}_{_\mathit{ R}} -\nabla_\mu
\nabla_\nu\mathit{ f}_{_\mathit{ R}}=-\frac{\kappa^2}{2}{{{\cal
T}^{{}^{{}^{^{}{\!\!\!\!\scriptstyle{em}}}}}}}_{\mu \nu} \,.
\end{eqnarray}

Consequently, Eqs.~\eqref{e1}, (\ref{f3}), and (\ref{f3ss}) must be examined under a spherically symmetric metric ansatz with two unknown functions in $(2+1)$ dimensions~\cite{Karakasis:2021lnq}.

%%%%%%%%%%%%%%%%%%%%%%%%%%%%%%%%%%% Section 3 %%%%%%%%%%%%%%%%%%%%%%%%%%%%%%%%%%%%%%%%
\section{The (2+1)-dimension  black hole solutions}\label{S3}
The metric of (2+1)-dimensional, expressed in $(t, r, \phi)$, takes the form:
\cite{Canate:2020btq}
\begin{eqnarray} \label{met12}
ds^2=-k(r)dt^2+\frac{dr^2}{k(r)k_1(r)}+r^2d\phi^2\,,
\end{eqnarray}
where $k(r)$ and $k_1(r)$ represent functions that depend on $r$. The Ricci scalar associated  Eq.~(\ref{met12}) is calculated to be:
  \begin{eqnarray} \label{Ricci}
  &&R(r)={\frac {1}{96{r}^{2}}}\left[4\,{r}^{2}k'' k_1 k' k'_1 +4\, k'  k_1  k'_1  k +4\,{k'_1}^{2}  k^{2 }-4\, {k'}^{2} k'_1rk_1 -8\, k''  k_1^{2}rk' -2\,k'{k'_1}^2rk -4\, k''k_1 rk'_1 k +{r}^{2} {k'}^2 {k'_1}^{2}+4\,{r}^{2} k''^2  k_1^{2}\right.\nonumber\\
  &&\left.+4\, {k'}^{2}  k_1^{2} \right] \,.
  \end{eqnarray}
Here, $k \equiv k(r)$ and $k_1 \equiv k_1(r)$ denote functions of the radial coordinate $r$, with $k' = \frac{dk}{dr}$ representing the first derivative of $k$ with respect to $r$, $k'' = \frac{d^2k}{dr^2}$ denoting the second derivative, and $k'_1 = \frac{dk_1}{dr}$ being the first derivative of $k_1$. By substituting Eqs. (\ref{f3}) and (\ref{f3ss}) into Eq. (\ref{met12}) and incorporating Eq. (\ref{Ricci}) into Eq.~(\ref{f3ss}), we get:
 \begin{eqnarray}\label{febtz}
&& {\mathop{\mathcal{ {\L}}}}_t{}^t=\frac {1}{12r}\left(2 \digamma  k' k_1-\digamma  k' k'_1 r-2\digamma  k'' k_1 r +4\digamma k'_1 k -2 \digamma' k'k_1 r+4\digamma''  k k_1 r+2 \digamma'  k'_1 k r+4 \digamma'  k k_1 -4 \xi'^{2}k_1 r\right)=0\,,\nonumber\\
&&{\mathop{\mathcal{{\L}}}}_r{}^r=-\frac {1}{12r}\left(2\digamma   k'' k_1  r+\digamma   k' k'_1 r-2\digamma   k' k_1  +2\digamma   k'_1 k +2  \digamma'  k'   k_1  r+ 8 \digamma''  k k_1  r+4 \digamma'    k'_1 k  r-4 \digamma'    k  k_1  +4\xi'^{2}k_1  r\right)=0\,,\nonumber\\
&&{\mathop{\mathcal{ {\L}}}}_\phi{}^\phi=-\frac {1}{6r}\left(2\digamma  k' k_1 +\digamma  k'_1 k -\digamma  k'  k'_1 r-2\digamma  k'' k_1 r-2  \digamma'  k'k_1 r- 2 \digamma'' k k_1 r- \digamma' k'_1 k r+4\digamma' kk_1 -4 \xi'^{2}k_1 r\right)=0\,,
\label{feq}
\end{eqnarray}
where $\xi\,\equiv \xi(r)$
is an-unknown { function} related to the electric which is defined from  vector potential
\begin{equation}\label{chr1}
A= \xi(r)dt\,,
\end{equation}
and $\xi'=\frac{d\xi}{dr}$. Using Eq. \eqref{met12}  in Eq. \eqref{e1} we get:
\begin{align}\label{ch}
\xi''r+\xi'=0\,.
\end{align}
Here, $\digamma \equiv \digamma(r) = \frac{df(R(r))}{dR(r)} = \frac{df(r)}{dr} \times \frac{dr}{dR}$, where $\digamma'$ and $\digamma''$ represent the first and second derivatives of $\digamma(r)$ with respect to $r$, respectively. Given the assumption of spherical symmetry for the spacetime, we posit that $f(R) = f(r)$. Ultimately,  the trace outlined in (\ref{f3}) assumes the following form:
\begin{align}
\label{trac}
&{\mathop{\mathcal{{\L}}}}=-\frac {1}{2r}\left(-4 \digamma'  k'k_1  r-4 \digamma''  k  k_1  r-2 \digamma'  kk'_1  r-4 \digamma' kk_1 +\digamma   k'k'_1 r+2 \digamma   k''k_1 r+4\digamma   k' k_1  +2\digamma  kk'_1  +3f  r\right)=0\,.
\end{align}
Next, we will examine specific instances of the aforementioned differential equations, as presented in Eqs. (\ref{febtz}), \eqref{ch}, and (\ref{trac}), with the aim of identifying analytical solutions\vspace{0.2cm}:
\subsection{The case: $\digamma(r)=a_0$ and    $k_1=1$, $\xi=0$}\footnote{It must be emphasized that $\digamma(r)\neq0$ in the present analysis, given that $\digamma(r)=0$ implies $f(R)=\text{constant}$, which lies outside the focus of our investigation.}
In the scenario where $\digamma(r)$ is a constant, denoted as $a_0$, and $k_1$ is set to 1, the differential equations (\ref{febtz}) simplify to the following differential equation:
\begin{align}
rk''-k'=0   \Rightarrow k=c_0+c_1r^2\,, \quad \mbox{where $c_0$ $\&$ $c_1$ denote constants.}\,\,\mbox{The form of k in this case is the BTZ form  \cite{Setare:2003hm}}.
\end{align}
A similar analysis and outcome can be applied when $\digamma(r)$ equals $a_0$ while $k_1$ is not equal to 1, and $\xi$ is zero.
%where $\Lambda$ and $m$ are integration constants.
\subsection{The case: $\digamma(r)=a_0$ and   $k_1=1$, $\xi\neq 0$}
With  $k_1=1$ and $\digamma(r)=a_0$  the differential equations, (\ref{febtz}) and \eqref{ch}, assume the following solution:
\begin{align}
&2r\xi'^2+ra_0k''-a_0k'=0, \quad \xi'+r\xi''=0, \quad \Rightarrow k=c_0+c_1r^2, \quad \xi=c_2+c_3ln\,r. \mbox{Here, we assign $a_0$ the value of 1, and } \nonumber\\
 &\mbox{ $c_2$ and $c_3$ represent constants. This solution conforms to the charged BTZ form \cite{Dehghani:2017thu,Nashed:2017fnd,Soroushfar:2015dfz}.}
\end{align}
\subsection{The case: $\digamma(r)\neq constant$,  $k_1=1$, and $\xi=0$}
In this case the solution of the resulting differential equation yields the form:
\begin{align}
\label{sp11}
& k(r)=\left[{c_5}^{2}-2c_5c_4r +2{c_4}^{2}\ln \left( c_4+\frac{c_5}r \right) {r}^{2} \right] c_0+{r}^{2}c_1, \qquad \digamma(r)=c_4 r+c_5\,.
\end{align}
When $c_4=0$ and $c_5=1$ we recover the  case {\bf A}.
\subsection{The case:  $\digamma(r)\neq constant$,  $k_1=1$, and $\xi\neq 0$}
In this scenario, the solution to the corresponding differential equations is as follows:
{
\begin{align}
\label{sp11c}
& k(r)=\left[ 2\,c_5\,c_4\,r-2\,{c_4}^{2}\ln  \left(\frac{ rc_4+c_5}{r} \right) {r}^{2}-{c_5}^{2} \right] c_0+{r}^{2}c_1 -\frac{{c_3}^{2}}{{ c_5}^{3}} \left[ 2\,{c_4}^{2}{ dilog} \left( { \frac {rc_5}{c_4}} \right) {r}^{2}+2\,{c_4}^{2} \ln  \left( rc_4+c_5 \right)\right.\nonumber\\
 &\left.\ln  \left( {\frac {rc_5}{c_4}} \right) {r}^{2}+2 {r}^{2}{c_4}^{2}\ln  \left( r \right) [1-\ln  \left( rc_4+c_5 \right)]+ \left( 2 \,c_5\,c_4\,r-{c_5}^{2} \right) \ln  \left( r \right) -\frac{{c_5}^{2}}2+2\,c_5\,c_4\,r \right], \nonumber\\
 & \digamma(r)=c_4 r+c_5\,, \qquad \xi=c_2+c_3ln\left(\frac{r}{r_0}\right)\,,
\end{align}

where $dilog$ is the Dilogarithm function}\footnote{The dilogarithm function is defined as follows: \[dilog(x)=\int_{t=1}^{t=x}\frac{\ln(t)}{1-t}dt\,,\] where $dilog(0)=\frac{\pi^2}{6}$, and $dilog(1)=0$.}. When $c_3=0$, we return to the scenario labeled as case {\bf C}. If both $c_3=c_4=0$, we revert to case {\bf A}, which corresponds to the BTZ solution. When $c_4=0$, we obtain case {\bf B}, representing the charged BTZ scenario.
\subsection{The case: $\digamma(r)\neq constant$, $k_1\neq 1$, and $\xi=0$}
When  $\digamma(r)\neq constant$ and when $k_1\neq 1$ we obtain:
\begin{align}
\label{sp111}
& k(r)=\frac{c_1r^6+4c_{0}c_6{}^2+6c_{0}c_6r^2+c_{0} r^4}{r^4}, \qquad k_1=\frac{e^{\frac{2c_6}{r^2}}}{(1+\frac{2c_6}{r^2})^4},\qquad \digamma(r)=-e^{-\frac{c_6}{r^2}}\,.
\end{align}
The uncharged BTZ scenario is retrieved when $c_6=0$.
\subsection{The case: $\digamma(r)\neq constant$,  $k_1\neq 1$, and $\xi\neq 0$}
Ultimately, for the scenario where  $\digamma(r)\neq constant$, $k_1\neq 1$ and $\xi\neq 0$ we obtain:
\begin{align}
\label{sp13d}
& k(r)={r}^{2}c_{1}+{\frac { \left( 4\,{c_6}^{3}+15\,r{c_6}^{2}+20\,{r}^{2}c_6 +{r}^{3} \right) c_{0}}{{r}^{3}}}-\frac{2{c_{7}}^{2}{{ e}^{-{\frac {c_6}{r} }}}}{{r}^{4}{c_6}^{2}} \left[ \frac{-{{ e}^{{\frac {c_6}{r}}}}r}5 \left( {c_6}^{3}+{\frac {15 }{4}}\,r{c_6}^{2}+5\,{r}^{2}c_6+\frac{5}2\,{r}^{3} \right){c_6}^{2}{ Ei} \left( 1,{\frac {c_6}{r}} \right)\right.\nonumber\\
& \left. +{c_6}^{6}+14\,r{c_6}^{5}+{\frac {8511}{ 20}}\,{r}^{3}{c_6}^{3}+{\frac {481}{5}}\,{r}^{2}{c_6}^{4}+{\frac {13109}{5 }}\,{r}^{5}c_6+{\frac {13109}{5}}\,{r}^{6}+{\frac {26133}{20}}\,{r}^{4}{ c_6}^{2} \right],\nonumber\\
& k_1=\frac{e^{\frac{2c_6}{r}}}{(1+\frac{c_6}{r})^{^6}}\,,\quad \digamma(r)=e^{-\frac{c_6}{r}}, \quad \xi=c_{2}+ \left[ {Ei} \left( 1,{\frac {c_6}{r}} \right) + \left( 8
\,{r}^{2}+5\,c_6r+{c_6}^{2} \right) \frac{{ e}^{-{\frac {c_6}{r}}}}{r^2}
 \right] c_{7}\,,
\end{align}
where $Ei$ is the exponential integrals\footnote{The exponential integrals, $Ei(a,x)$ are defined for $Re(x) > 0$ by:\\
\[Ei(a,x)=\int_1^\infty e^{-bx}b^{-a} db.\]}.
%If $c_3=0$ we revert to case  {\bf E}.
If    $c_3=c_6=0$ we return to  the case, {\bf A} corresponding to the uncharged BTZ scenario. Setting $c_6=0$ leads us to case {\bf B}, which represents the charged BTZ case.
In the following section, we will delve into the physical implications of each solution, excluding the first two cases since they are well-established solutions within GR (GR).
\section{Analysis of the Physical Properties of the black hole Solutions}\label{BH}
Now let us study each case of the pervious black hole solutions, excluding cases {\bf A} and {\bf B}, in details:
\subsection{The case $\digamma(r)\neq constant$ and   $k_1=1$, $\xi=0$}
 In this case, the line-element takes the form:
\begin{align}
\label{line1}
&ds^2=-\left[\left\{{c_5}^{2}-2c_5c_4r +2{c_4}^{2}\ln \left(\frac{ c_4r+c_5}r \right) {r}^{2} \right\} c_0+{r}^{2}c_1\right]dt^2+\frac{dr^2}{\left\{{c_5}^{2}-2c_5c_4r +2{c_4}^{2}\ln \left(\frac{ c_4r+c_5}r\right) {r}^{2} \right\} c_0+{r}^{2}c_1 }+r^2d\phi^2\,.
\end{align}
{For large and small values of $r$, the line element takes the asymptotic forms
 \begin{align}
\label{line1as}
&ds^2_{r\to \infty}\approx-\left[\Lambda_{eff} {r}^{2}+\frac{2}{3}{\frac {mc_5^3}{c_4r}}-\frac{1}{2}{
\frac {mc_5^4}{{c_4^2r}^{2}}}+\frac{2}5{\frac {
mc_5^5}{c_4^3{r}^{3}}}
\right]dt^2+\frac{dr^2}{\left[\Lambda_{eff} {r}^{2}+\frac{2}{3}{\frac {mc_5^3}{c_4r}}-\frac{1}{2}{
\frac {mc_5^4}{{c_4^2r}^{2}}}+\frac{2}5{\frac {
mc_5^5}{c_4^3{r}^{3}}}\right]}+r^2d\phi^2\,,\nonumber\\
&ds^2_{r\to 0}\approx-\left[mc_5^2-2\,m\,rc_4c_5+ {\Lambda_1}_{eff} {r}^{2}+\frac{2m {r}^{3}{c_4}^{3}}{c_5}\right]dt^2+\frac{dr^2}{mc_5^2-2\,m\,rc_4c_5+ {\Lambda_1}_{eff} {r}^{2}+\frac{2m {r}^{3}{c_4}^{3}}{c_5}}+r^2d\phi^2\,.
\end{align}
where the identification $c_0=m$ has been made, and
 \begin{align}\label{lam}
 {\Lambda}_{eff}=\Lambda-2c_4^2c_0\ln c_4,   \quad \mbox{and} \quad {\Lambda_1}_{eff}=\Lambda  +2c_4^2m\,\ln
 \left( {\frac{r} {c_5}} \right), \quad\mbox{with}\quad  \Lambda=c_1.\end{align}  

When $c_4=0$ and $c_5=1$, the above expressions reduce to the BTZ solution. From Eq.~\eqref{line1as}, however, it is evident that at large $r$ the parameter $c_4$ cannot vanish, which implies that the solution does not simply collapse into the BTZ form of GR. A deeper investigation of the large-$r$ regime is therefore required and will be addressed separately.  

The asymptotic properties of the metric function deserve careful discussion. At short distances, the solution approaches a BTZ-like black hole, but with modifications: higher-curvature contributions manifest themselves through the effective cosmological constant ${\Lambda_1}_{eff}$. Additional corrections to the metric appear as ${\cal O}(r^{n})$ terms (for $n\geq 1$), and these corrections are entirely governed by $c_4$, which reflects the influence of higher-order curvature. The small-$r$ expansion shows qualitative differences from the usual BTZ black hole: in addition to constant and ${\cal O}(r^2)$ terms, there are linear contributions and logarithmic corrections of the form ${\cal O}(r^2\ln r)$, which significantly affect the near-origin behavior. The behavior of the metric \eqref{line1} in these limits is illustrated in Fig.~\ref{Fig:11}.  

\begin{figure}
\centering
\subfigure[~Behavior of Eq.~\eqref{line1as} as $r\to \infty$]{\label{fig:g00i}\includegraphics[scale=0.3]{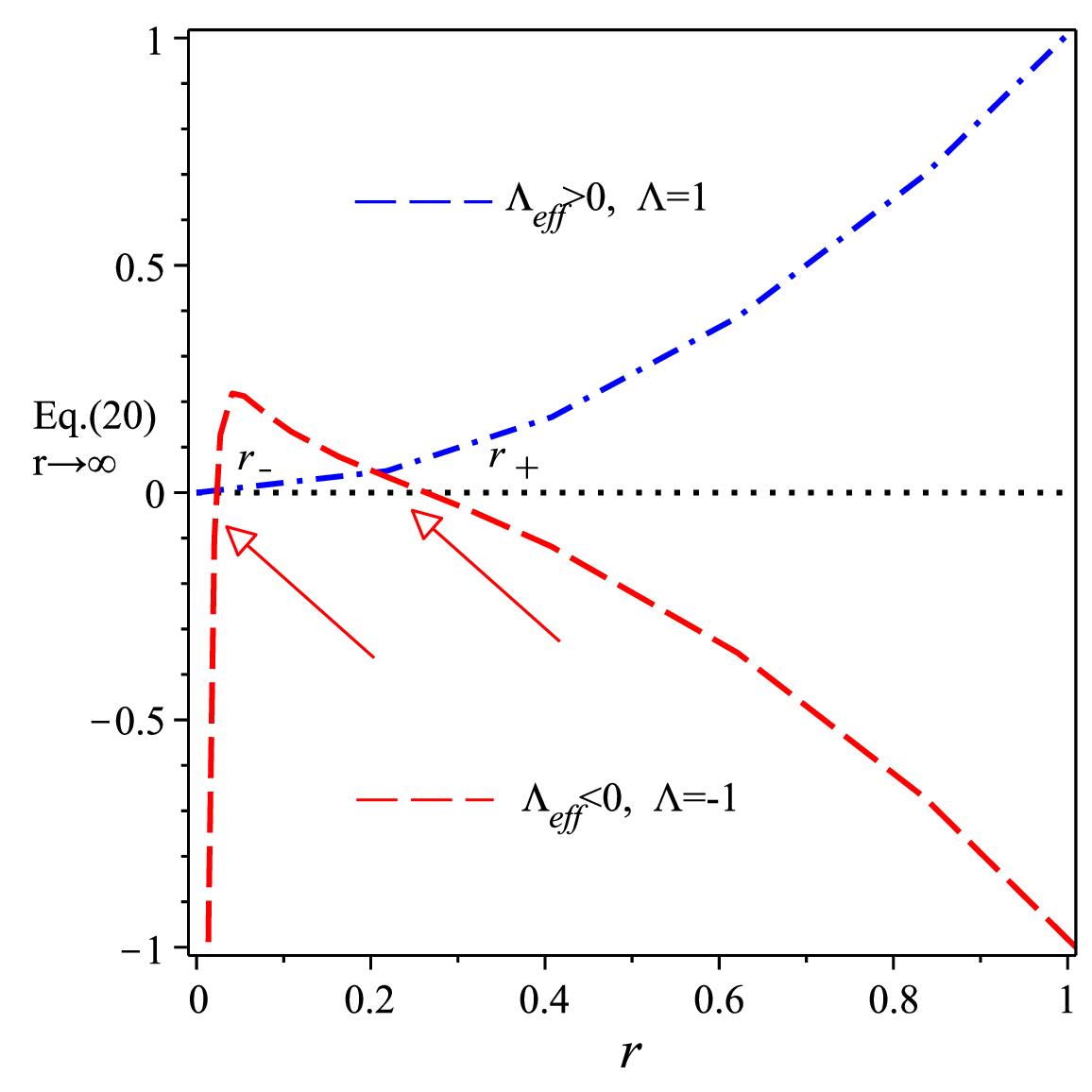}}
\subfigure[~Behavior of Eq.~\eqref{line1as} as $r\to 0$]{\label{fig:goo0}\includegraphics[scale=0.3]{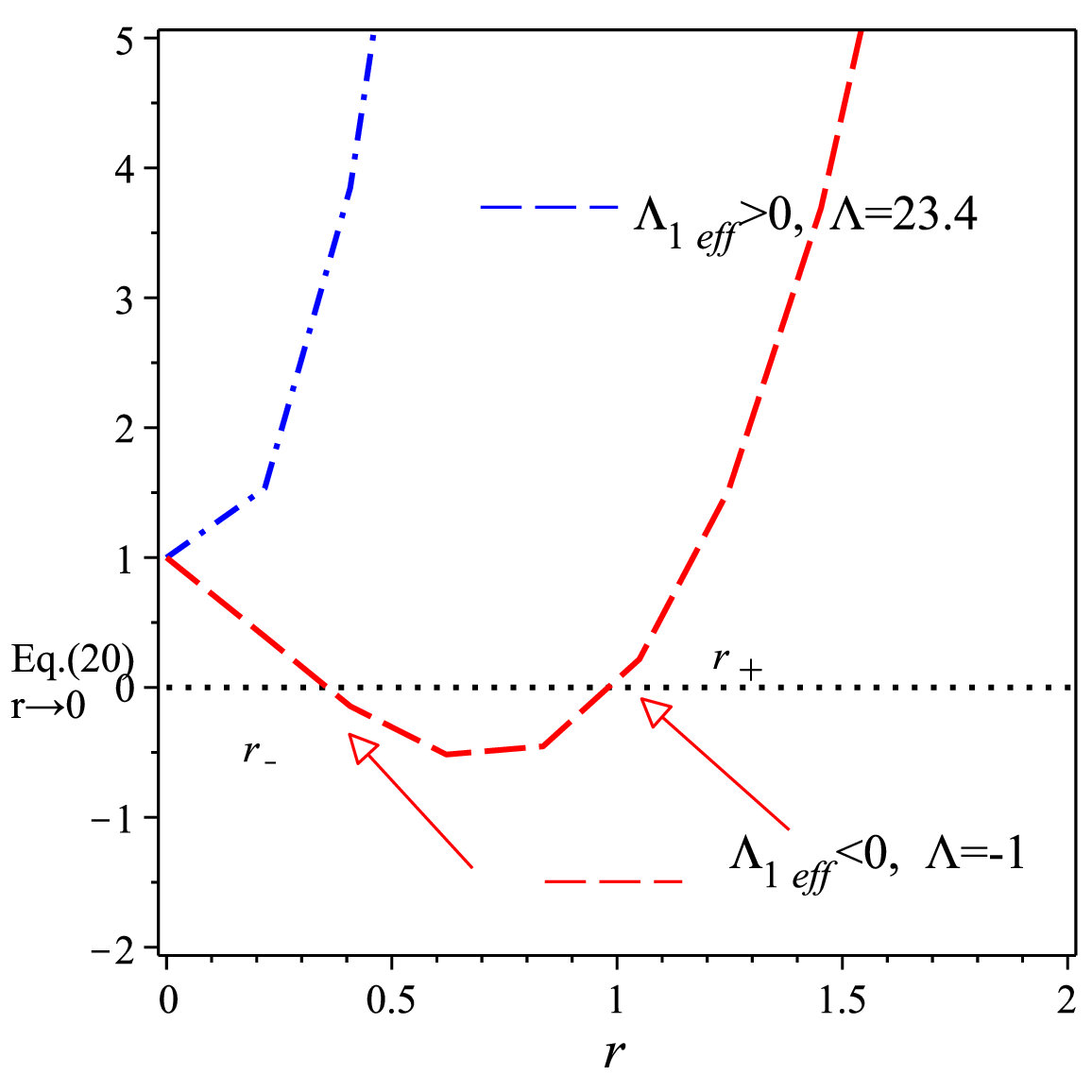}}
\caption[figtopcap]{\small{Plots showing the asymptotic behavior of the metric function: 
\subref{fig:g00i} for $r\to \infty$; 
\subref{fig:goo0} for $r\to 0$. 
The curves are obtained using $M=0.3$ and $c_2=c_3=c_4=c_5=c_6=1$.}}
\label{Fig:11}
\end{figure}

Equation~(\ref{line1}) clearly demonstrates that an effective cosmological constant arises naturally, even without introducing it by hand.  

The curvature invariants corresponding to the line element \eqref{line1} can be evaluated as
\begin{align}\label{invI}
&(R_{\mu \nu \alpha \beta}R^{\mu \nu \alpha \beta}=R_{\mu \nu}R^{\mu \nu})_{r \to \infty}\approx 12\Lambda_{eff}^2-\frac{4m\Lambda_{eff} }{c_4^2r^4}+\frac{48m\Lambda_{eff} }{5c_4^3r^5}\,,\quad (R_{\mu \nu \alpha \beta}R^{\mu \nu \alpha \beta})_{r \to 0}\approx C_0+C_1\,r+ C_2r^2\,,\nonumber\\
& (R_{\mu \nu}R^{\mu \nu})_{r \to 0}\approx C_3+C_4\,r+C_5\,r^2\,, \quad (R)_{r \to \infty} \approx \frac{m^2}{6c_4^2r^6}-\frac{2m^2}{3c_4^3r^7}+\frac{5m^2}{3c_4^4r^8}\,,\quad (R)_{r \to 0} \approx C_6+C_7\,r+C_8\,r^2\,.\nonumber\\
\end{align}
The coefficients $C_i$ $(i=0,\dots,8)$ are functions of the parameters $m$, $\Lambda_{eff}$, ${\Lambda_1}_{eff}$, and $c_4$. The quantities $\big(R_{\mu \nu \rho \sigma}R^{\mu \nu \rho \sigma},\,R_{\mu \nu}R^{\mu \nu},\,R\big)$ represent, respectively, the Kretschmann scalar, the Ricci tensor squared, and the Ricci scalar.

As a consistency check, by setting $c_4=0$ and $c_5=1$ in Eq.~\eqref{line1}, one indeed recovers the BTZ black hole solution~\cite{Banados:1992wn} with $k(r)=m+\Lambda r^2$. Thus, in the absence of higher-curvature corrections ($c_4=0$), GR is recovered. Accordingly, the metric \eqref{line1} can be interpreted as a generalized BTZ black hole in modified gravity. For $r\to 0$, the invariants revert to those of GR in the case $c_4=0$. However, the limit $r\to \infty$ for $c_4=0$ requires additional analysis, and this scenario will be explored further in future work.
}

Finally, using the trace of  Eq. \eqref{trac} we get:
\begin{align}\label{frI}
f(r)=-{\frac {8m\ln  \left( 1+c_5/r \right) {r}^{2}+8m\ln  \left( 1+c_5/r \right) rc_5 -8mc_5r-4m{c_5}^{2}-\Lambda{c_5}^{2}{r} ^{2}-\Lambda{c_5}^{3}r}{{c_5}^{2}r \left( r+c_5 \right) }}\,.
\end{align}
If we use Eq. \eqref{invI} to derive $r(R)$ we get:
\begin{align}\label{rR}
r=\pm\frac{1}{2}{\frac {C_7+\sqrt {{C_7}^{2}+4\,C_8\,R-4\,
C_8\,C_6}}{C_8}}\,.
\end{align}
Using Eq. \eqref{rR} in Eq. \eqref{frI} we get $f(R)$ as:
\begin{align}\label{fr1}
f(R)\approx C_{9}+C_{10}R+C_{11}R^2\,,
\end{align}
where $C_i, \, i=9\cdots 11$ are lengthy constants which depend on $m$, $\Lambda$ and $c_5$. Equation \eqref{fr1} shows that the form of $f(R)$ of the case {\bf C} is a polynomial function.   { So far we have not imposed any condition on $c_1=\Lambda$, therefore the spacetime might be asymptotically AdS or dS depending on the value of parameter

\begin{align}\label{cons} \nonumber\\
&\Lambda>2c_4^2c_0\ln c_4>0, \quad \mbox{asymptotically AdS}, \qquad \Lambda<2c_4^2c_0\ln c_4<0, \quad \mbox{asymptotically dS}, \quad \mbox{for} \quad r\to \infty\nonumber\\
&\Lambda<-2c_4^2m\,\ln
 \left( {\frac{r} {c_5}} \right), \quad \mbox{asymptotically AdS}, \qquad \Lambda>-2c_4^2m\,\ln
 \left( {\frac{r} {c_5}} \right), \quad \mbox{asymptotically dS}, \quad \mbox{for} \quad r\to 0.\end{align} We have shown the  behavior of the line-element \eqref{line1as}   in Fig.~\ref{Fig:11} using the above constrains presented in Eq.~\eqref{cons}. }

 { Before we close this subsection, it is important to stress that the problem of generalizing the
Schwarzschild-(anti)de Sitter solution in the case of a single metric degree of freedom using four dimensional spacetime has been discussed in \cite{Laporte:2024oaf}. In this study, \cite{Laporte:2024oaf},  the authors have shown that their results do not favor $f(R)$ gravitational models with a single metric unlike the case of three dimension which is supported by Eq.~\eqref{line1}.}
\subsection{The case $\digamma(r)\neq constant$ and   $k_1=1$, $\xi\neq0$}
 In this case, the line-element takes the form:
\begin{align}
\label{line2}
&ds^2=-k(r)dt^2+\frac{dr^2}{k(r)}+r^2d\phi^2\,,
\end{align}
where $k(r)$ is defined in Eq. \eqref{sp11c}.
{ The above line element behaves for large/small $r$ as:
  \begin{align}
\label{line2as1}
&ds^2_{r\to \infty}\approx -\left[{\Lambda_2}_{eff}\, {r}^{2}+C_{12}r+C_{13}+{\frac {C_{14}}{4r}}-{
\frac {C_{15}}{{r}^{2}}}
\right]dt^2+\frac{dr^2}{\left[{\Lambda_2}_{eff}\, {r}^{2}+C_{12}r+C_{13}+{\frac {C_{14}}{4r}}-{
\frac {C_{15}}{{r}^{2}}}\right]}+r^2d\phi^2\,,\nonumber\\
&ds^2_{r\to 0}\approx-\left[C_{16}+C_{17}r+{\Lambda_3}_{eff}r^2+C_{18}r^3\right]dt^2+\frac{dr^2}{C_{16}+C_{17}r+{\Lambda_3}_{eff}r^2+C_{18}r^3}+r^2d\phi^2\,.
\end{align}
where  ${\Lambda_2}_{eff}$ and ${\Lambda_3}_{eff}$ are effective cosmological constants for large/small $r$ and $C_{12}\cdots C_{18}$ are  depend mainly on $c_4$, $c_5$ and $c_3$\footnote{{ The form of the effective cosmological constants ${\Lambda_2}_{eff}$ and ${\Lambda_3}_{eff}$ are given as ${\Lambda_2}_{eff}=-\frac{1}{3c_5^3}[6c_4^2c_3^2(\ln c_5-2\ln c_3\ln c_5+3\ln c_3)-6c_3^2c_5^3\ln c_3+3c_5^3\Lambda+\pi^2c_3^2c_4^2],$
 ${\Lambda_3}_{eff}=-\frac{2}{c_5^3}[6c_4^2c_3^2(2\ln c_5+\ln r-\ln c_4\ln c_5)-6c_4^2c_5^3\ln(r/c_5)-3\Lambda c_5^3+\pi^2c_3^2c_4^2].$}}.  The BTZ solution can be recovered if $c_4=c_3=0$. The case $\digamma(r)\neq constant$ and   $k_1=1$, $\xi=0$ can be recover when $c_3=0$. For large/small  $r$ the line element \eqref{line2as1} is different from the charged BTZ due to the contribution of the higher order curvature. Following the procedure done for the case $\digamma(r)\neq constant$ and   $k_1=1$, $\xi=0$  we can show that for the effective cosmological constant for small $r$ we can create AdS spacetime with two horizons one of them representing the $r_+$.

 The  invariants of the line-element \eqref{line2as1} yields:
\begin{align}\label{inv}
&(R_{\mu \nu \alpha \beta}R^{\mu \nu \alpha \beta}=R_{\mu \nu}R^{\mu \nu})_{r \to \infty}\approx C_{19}+\frac{C_{20}}{r}+\frac{C_{21}}{r^2}\,, \qquad (R_{\mu \nu \alpha \beta}R^{\mu \nu \alpha \beta}=R_{\mu \nu}R^{\mu \nu})_{r \to 0}\approx \frac{C_{22}}{r}+\frac{C_{23}}{r^2} \,,\nonumber\\
&(R)_{r \to 0} \approx C_{24}+\frac{C_{25}}{r}\,, \qquad (R)_{r \to \infty} \approx C_{26}+\frac{C_{27}}{r}\,,
\end{align}
where $C_{19}\cdots C_{27}$ are constants depend mainly on $c_4$, $c_5$ and $c_3$.  The constants $C_{20}\cdots C_{23}$ and $C_{25}\cdots C_{27}$ are vanishing when $c_4=c_3=0$ and $c_5=1$ and this yields $C_{19}=C_{24}\equiv \Lambda$ which is  the BTZ. The black hole of this case is singular as we can see that the invariants are singular as $r \to 0$. This case reduce to the charged GR black hole as the constant $c_4=0$ and to the uncharged when $c_4=c_3=0$.}

Finally, using  Eq. \eqref{trac} we can calculate $f(r)$ from which if we follow the same procedure done in the case $\digamma(r)\neq constant$ and   $k_1=1$, $\xi\neq 0$ we can show that the $f(R)$  is a polynomial function.
%\begin{align}\label{fr}
%f(r)=-\frac{4\, \left( 16\,m\,{c_6}^{4}+24\,m\,{r}^{2}{c_6}^{3}+16\,m\,{r}^{4}{c_6}^{2}-8\,\Lambda\,{r}^{6}
%{c_6}^{2}+12\,m\,{r}^{6}c_6+6\,\Lambda\,{r}^{8}c_8+\Lambda_1\,{
%r}^{10} \right) {e^{{\frac {c_6}{{r}^{2}}}}}}{\left( 2\,c_6+{r}^{2} \right) ^{5}}+c_{9}\,.
%\end{align}
%If we use the same procedure done in the case $\digamma(r)\neq constant$ and   $k_1=1$, $\xi=0$ we can show that the $f(R)$  is a polynomial function.
 \subsection{The case $\digamma(r)\neq constant$ and   $k_1\neq 1$, $\xi=0$}
{ In this case, the line-element takes the form:
\begin{align}
\label{line3}
&ds^2=-k(r)dt^2+\frac{dr^2}{k(r)k_1(r)}+r^2d\phi^2\,,
\end{align}
where $k(r)$ and $k_1(r)$ are defined in Eq. \eqref{sp11c}.
The above line element behaves for large/small $r$ as:
%\footnote{Due to the existence of exponential term in line-element \eqref{line3}, therefore we cannot  asymptote this metric for small $r$.}:
 \begin{align}
\label{line3as}
&ds^2_{r\to \infty}\approx-\left[\Lambda r^2+m+\frac{2mc_6}{r^2}+\frac{2mc_6^2}{3r^4}\right]dt^2+\frac{dr^2}{\Lambda{r}^{2}+C_{28}+\frac{C_{29}}{r^2}+\frac{C_{30}}{r^4}}+r^2d\phi^2\,,\nonumber\\
&ds^2_{r\to 0}\approx-\left[\Lambda r^2+m+\frac{2mc_6}{r^2}+\frac{2mc_6^2}{3r^4}\right]dt^2+\frac{dr^2}{\Lambda{r}^{2}+m+\frac{mr^2}{4c_6}+\frac{mr^4}{8c_6^2}}+r^2d\phi^2\,,
\end{align}
where  $C_{28} \cdots C_{30}$  are constants depend mainly on $c_6$ and they are vanishing if $c_6=0$.  The BTZ solution can be discovered if $c_6=0$ for $r \to \infty$ however, for $r\to 0$ the constant $c_6$ is not allowed. This means that solution presented in Eq. \eqref{sp11c} need more investigation for $r\to 0$.
 The  invariants of the line-element \eqref{line3as} yields:
\begin{align}\label{inv1}
&(R_{\mu \nu \alpha \beta}R^{\mu \nu \alpha \beta}=R_{\mu \nu}R^{\mu \nu})_{r \to \infty}\approx 12\Lambda^2-\frac{48\Lambda_1^2c_6}{r^2}+\frac{C_{31}}{r^4}\,, \qquad (R)_{r \to \infty} \approx \frac{6\Lambda^2c_6^2}{r^4}+\frac{C_{32}}{r^6}\,,\nonumber\\
&(R_{\mu \nu \alpha \beta}R^{\mu \nu \alpha \beta}=R_{\mu \nu}R^{\mu \nu})_{r \to 0}\approx C_{33}r^4+C_{34}r^2+\frac{131m^2}{4c_6^2}-\frac{32m^2}{c_6r^2}+\frac{40m^2}{r^4}\,, \qquad (R)_{r \to 0} \approx C_{35}r^2+\frac{5m^2}{3c_6^2}-\frac{7m^2}{4c_6r^2}\,,
\end{align}
where $C_{31}$ and $C_{32}$ are constant depend on $c_6$ and vanishing when $c_6=0$ and in that case the invaginates as $r \to \infty$ are consistent with those of GR. Also the constants  $C_{33}$ and $C_{34}$ are depend on $c_1=\Lambda$ and $c_0=m$.  In the case $r \to 0$ we see that the constant $c_6$ should not equal to zero.  The case $c_6\neq 0$ for $r\to 0$ needs more study. As we see that the case when $c_6=0$ has an issue when $r\to 0$.}
  %These constants are vanishing when $c_6$ equal constant and in that case we recover the BTZ.
Finally, using  Eq. \eqref{trac} we get the form of $f(r)$ in the form:
\begin{align}\label{fr}
f(r)=-\frac{4\, \left( 16\,m\,{c_6}^{4}+24\,m\,{r}^{2}{c_6}^{3}+16\,m\,{r}^{4}{c_6}^{2}-8\,\Lambda\,{r}^{6}
{c_6}^{2}+12\,m\,{r}^{6}c_6+6\,\Lambda\,{r}^{8}c_6+\Lambda_1\,{
r}^{10} \right) {e^{{\frac {c_6}{{r}^{2}}}}}}{\left( 2
\,c_6+{r}^{2} \right) ^{5}}+c_{9}\,.
\end{align}
If we use the same procedure done in the case $\digamma(r)\neq constant$ and   $k_1=1$, $\xi=0$ we can show that the $f(R)$  is a polynomial function.

\subsection{The case $\digamma(r)\neq constant$ and   $k_1\neq1$, $\xi\neq0$}
 In this case, the line-element takes the form:
\begin{align}
\label{line4}
&ds^2=-k(r)dt^2+\frac{dr^2}{k(r)k_1(r)}+r^2d\phi^2\,,
\end{align}
where $k$ and $k_1$ are defined in Eq. \eqref{sp13d}.
The above line element behaves for large $r$ as:
 \begin{align}
\label{line3as}
 &ds^2_{r\to \infty}=-\left[\Lambda_2 r^2+C_{35}+\frac{C_{36}}r-\frac{C_{37}}{r^2}\right]dt^2+\frac{dr^2}{C_{38}r^2+\frac{c_3^2r}{5c_6}+C_{39}+\frac{C_{40}}{r} }+r^2d\phi^2\,,\nonumber\\
 &ds^2_{r\to 0}=-\left[\Lambda_2 r^2+C_{35}+\frac{C_{36}}r-\frac{C_{37}}{r^2}\right]dt^2+\frac{1}{c_6^6}\frac{dr^2}{C_{41}+C_{42}r+C_{43}r^2 +C_{44}r^3 +C_{45}r^4}+r^2d\phi^2\,,\nonumber\\
\end{align}
where  $\Lambda_2$ depends on $c_{1}$ and $c_{5}$ and $C_{35} \cdots C_{45}$  are constants depend mainly on $c_0$, $c_{6}$, $c_3$ and  $\gamma$ which is the  Euler-Mascheroni.\footnote{Euler-Mascheroni is a constant that appears from the asymptote of the ${ Ei} \left( 1,{\frac {c_6}{r}} \right)$. which is the exponential integrals.}   The BTZ solution can be discovered if $c_{3}$ and $c_{6}=0$.
 The  invariants of the line-element \eqref{line3as} yields
 %\footnote{ {It is important to mention that the asymptote of the invariants for small $r$ in the cases that have $k_1(r)\neq 1$,  and   $\digamma(r)=\e^{-\frac{c_6}{r^2}}$ cannot be carried out because of the existence of this exponential term.}}:
 {
\begin{align}\label{inv2}
&(R_{\mu \nu \alpha \beta}R^{\mu \nu \alpha \beta}=R_{\mu \nu}R^{\mu \nu})_{r \to \infty}\approx 12\Lambda_2-\frac{C_{46}}{r}+\frac{C_{47}}{r^2}\,, \qquad (R)_{r \to \infty} \approx \frac{C_{48}}{r^2}+\frac{C_{49}}{r^3}\,,\nonumber\\
&(R_{\mu \nu \alpha \beta}R^{\mu \nu \alpha \beta}=R_{\mu \nu}R^{\mu \nu})_{r \to 0}\approx C_{49}+C_{50}r+C_{51}r^2\,, \qquad (R)_{r \to 0} \approx C_{52}+C_{53}r+C_{54}r^2\,,\nonumber\\
\end{align}
where the $C_{46} \cdots C_{54}$  are constants depend mainly on $c_{6}$ and $c_3$.  When $c_6=0$, the constants $C_{49} \cdots C_{54}$ become undefined because their definitions have $c_6$ on dominator. The constants $C_{46} \cdots C_{48}$ are vanishing when $c_{3}$ and  $c_{6}$ are vanishing and in that case we recover the BTZ.  }

Finally, using  Eq. \eqref{trac} we get:
\begin{align}\label{fr}
&f(r)=\int \frac{16}{{c_6}{r}^{3}\left( r+c_6 \right) ^{8} } \left[ \frac{2}{15} \left[ {c_6}^{6}+15{r}^{6}+14{r}^{2}{c_6}^{4} +{\frac {49}{2}}{r}^{3}{c_6}^{3}+25{r}^{4}{c_6}^{2}+15{r}^{5}c_6+5r{ c_6}^{5} \right] {e^{{\frac {c_6}{r}}}}r{c_6}^{2}{c_3}^{2}{ Ei} \left( 1,{\frac {c_6}{r}} \right)\right.\nonumber\\
&\left. +\frac{4}3 \left[ {c_6}^{6}c_0+5 c_0r{c_6}^{5}+14c_0{r}^{2}{c_6}^{4}+ \left({\frac {49}{2}}{r}^{3}c_0 -{r}^{5} c_1 \right) {c_6}^{3}+5{r}^{ 4} \left( 5c_0+{r}^{2}c_1 \right) {c_6}^{2}+ \left( 15{r}^{5}c_0-\frac{13} 2{r}^{7}c_1 \right) c_6+15{r}^{6}c_0-2{r}^{8}c_1 \right]\right.\nonumber\\
&\left.  r{c_6}^{2}{e^{{\frac {c_6}{r}}}}+ \left( {c_6}^{9}+{\frac {209744}{15}}{r}^{9}+{\frac {818}{15}}{r}^{ 2}{c_6}^{7}+{\frac {504}{5}}{c_6}^{6}{r}^{3}-{\frac {11563}{5}}{r}^{5} {c_6}^{4}-{\frac {1564}{5}}{r}^{4}{c_6}^{5}+{\frac {52486}{3}}{r}^{7}{ c_6}^{2}+{\frac {297139}{5}}{r}^{8}c_6\right.\right.\nonumber\\
&\left.\left. -{\frac {8677}{3}}{r}^{6}{c_6}^{3} +{\frac {34}{3}}r{c_6}^{8} \right) {c_3}^{2} \right] {dr}+c_{9}\,.
\end{align}
If we use the same procedure done in the case $\digamma(r)\neq constant$ and   $k_1=1$, $\xi=0$ we can show that the $f(R)$  is a polynomial function.

\begin{figure}
\centering
\subfigure[~The properties exhibited by $R$, $f(r)$, $f_R$, and $f_{RR}$ of Eq.~\eqref{sp11} $\sim$ \eqref{sp13d}]{\label{fig:R}\includegraphics[scale=0.24]{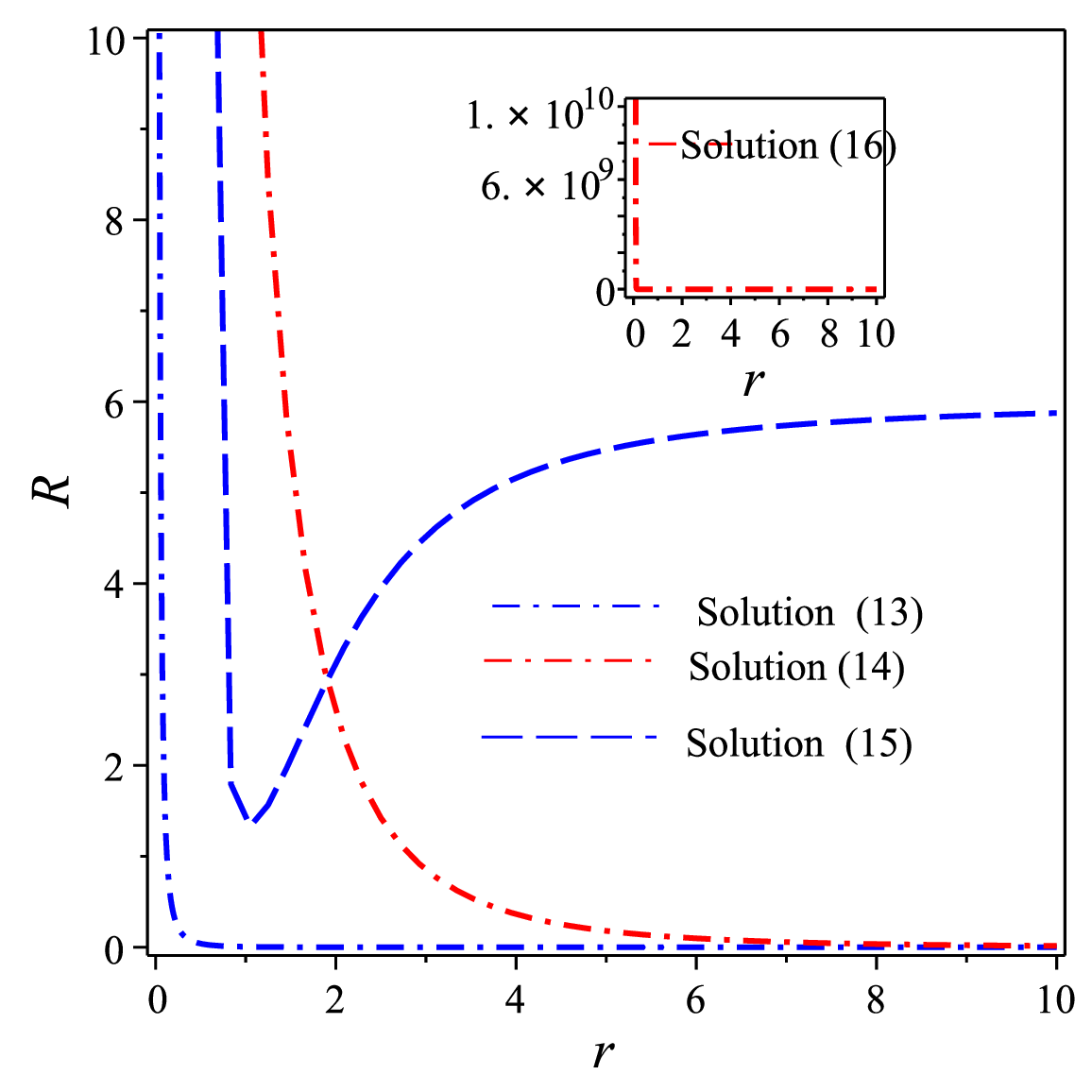}}
\subfigure[~The properties exhibited by $f(r)$ of   Eqs. \eqref{sp11} $\&$ \eqref{sp11c}]{\label{fig:fr}\includegraphics[scale=0.24]{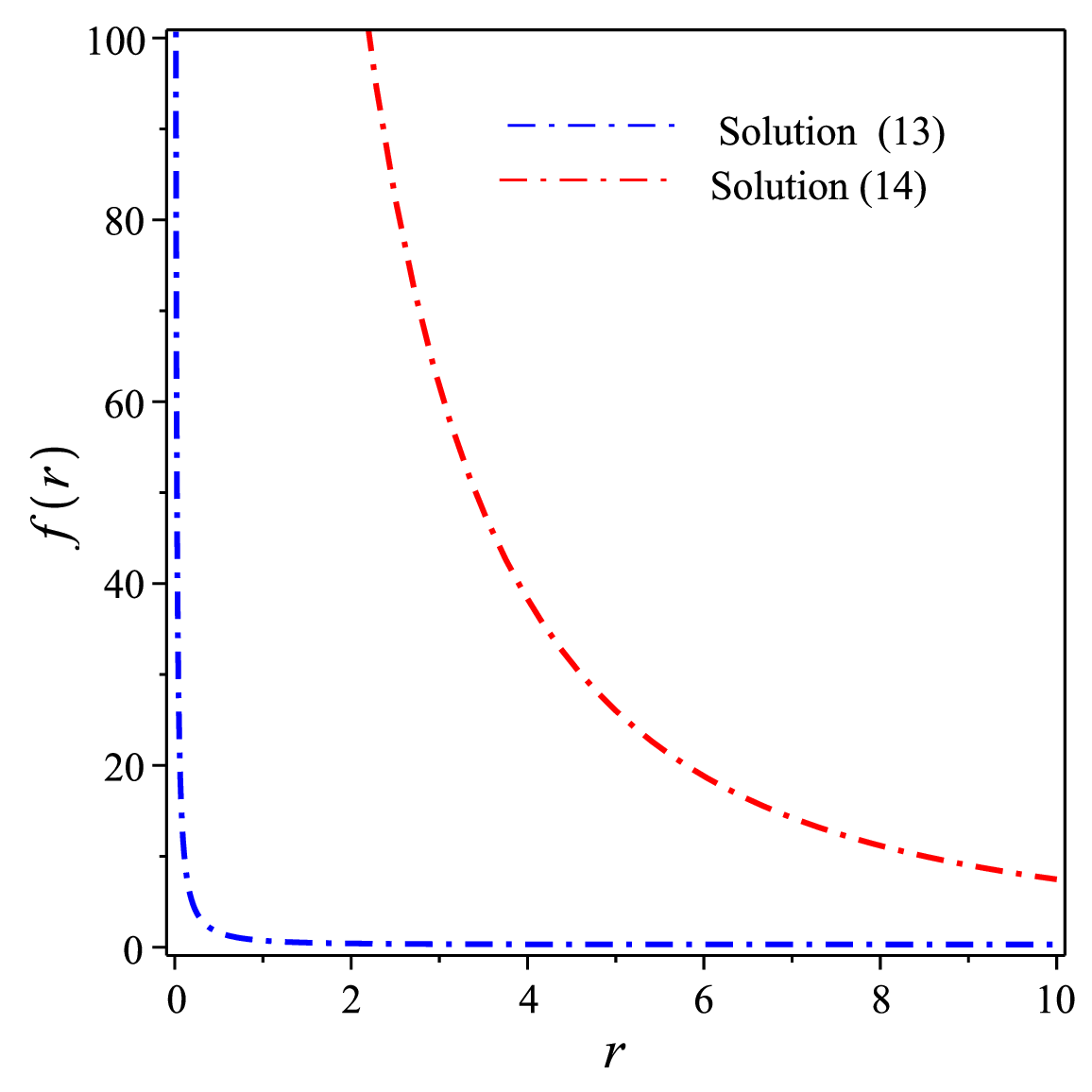}}
\subfigure[~The properties exhibited by $f(r)$ of  Eqs. \eqref{sp111} $\&$ \eqref{sp13d}]{\label{fig:fr1}\includegraphics[scale=0.24]{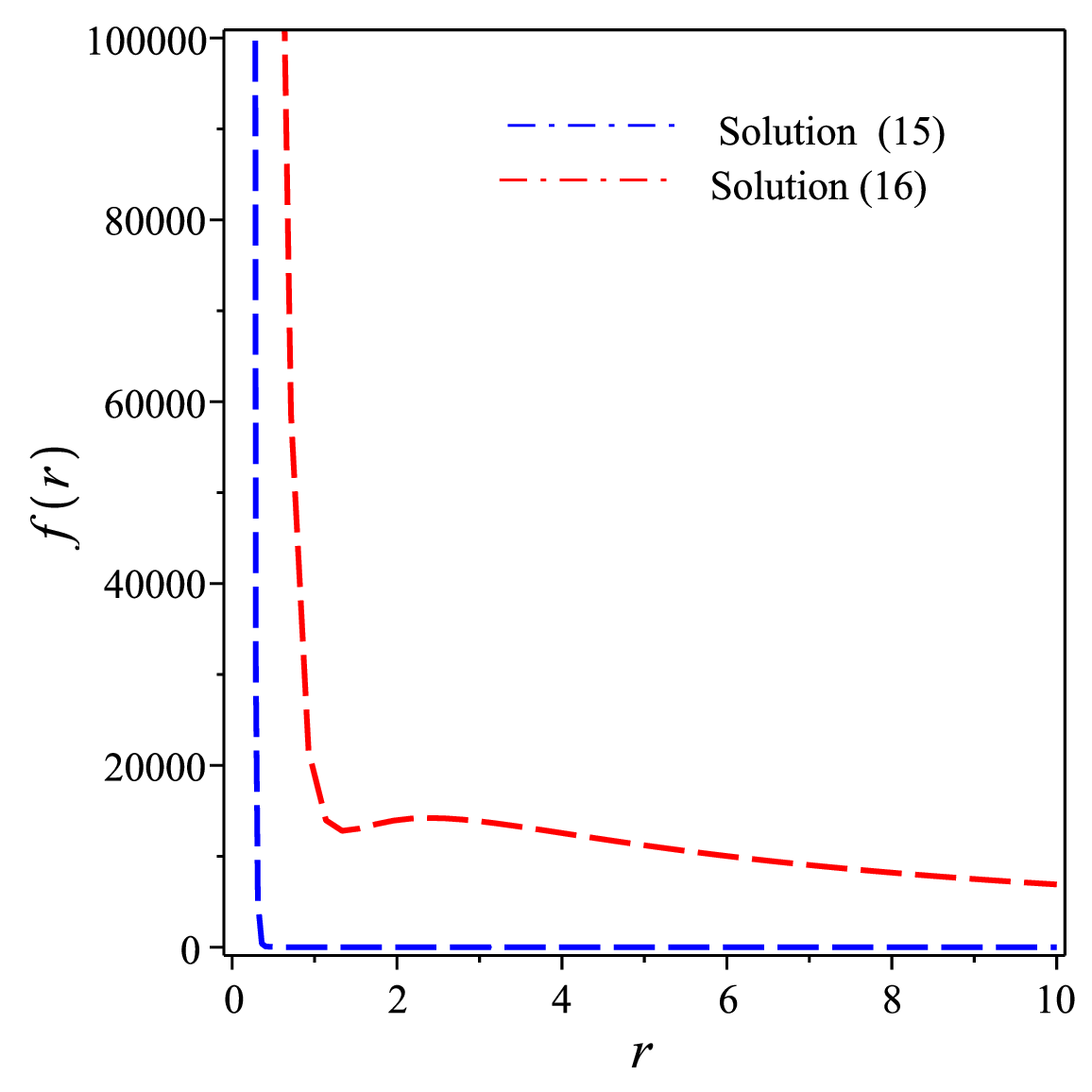}}
\subfigure[~The properties exhibited by  $f_R$ of  Eqs. \eqref{sp11} $\sim$ \eqref{sp13d}]{\label{fig:fR}\includegraphics[scale=0.24]{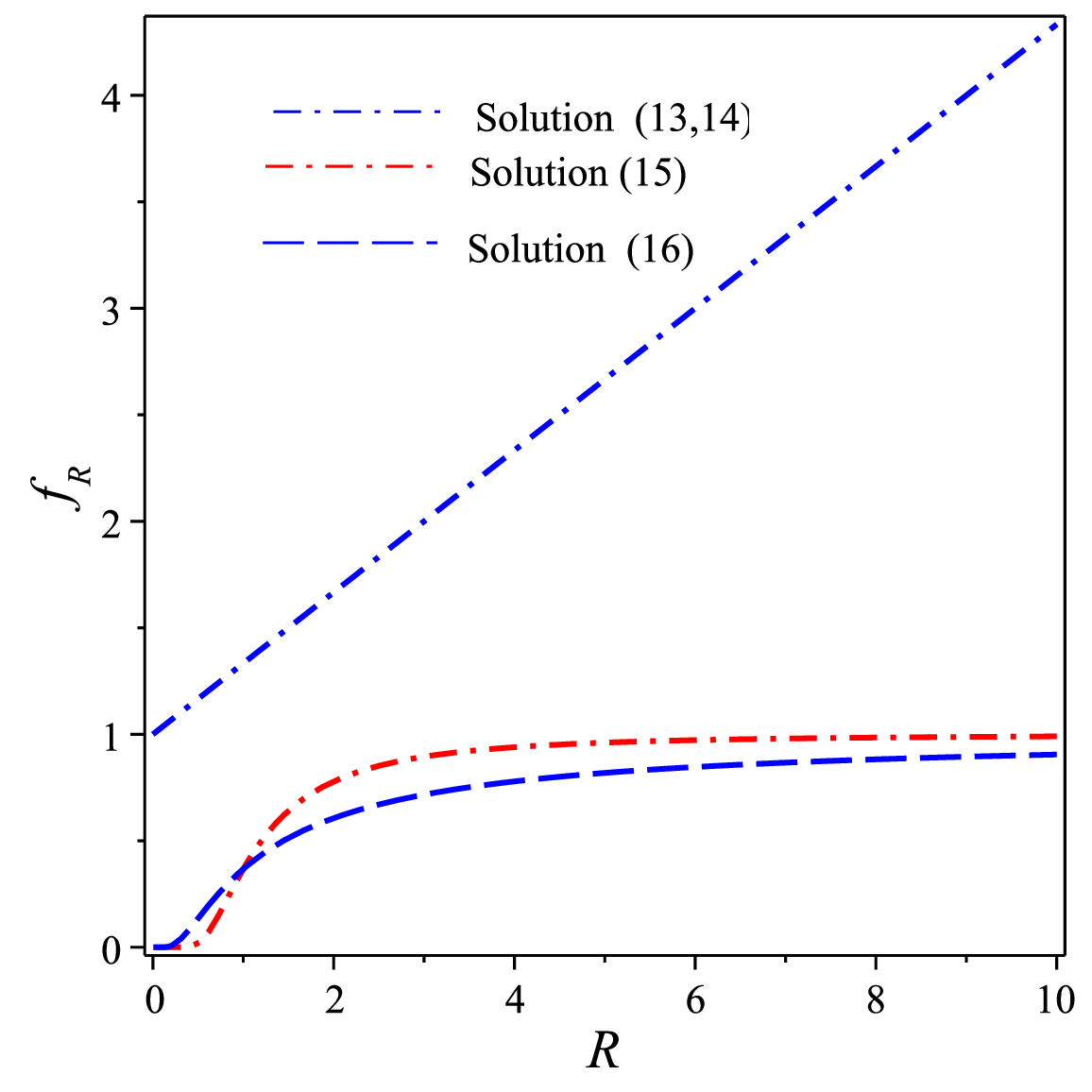}}
\subfigure[~The behavior of the function $f_{RR}$ of the solutions given by Eqs. \eqref{sp11} $\&$ \eqref{sp11c}]{\label{fig:fRR}\includegraphics[scale=0.24]{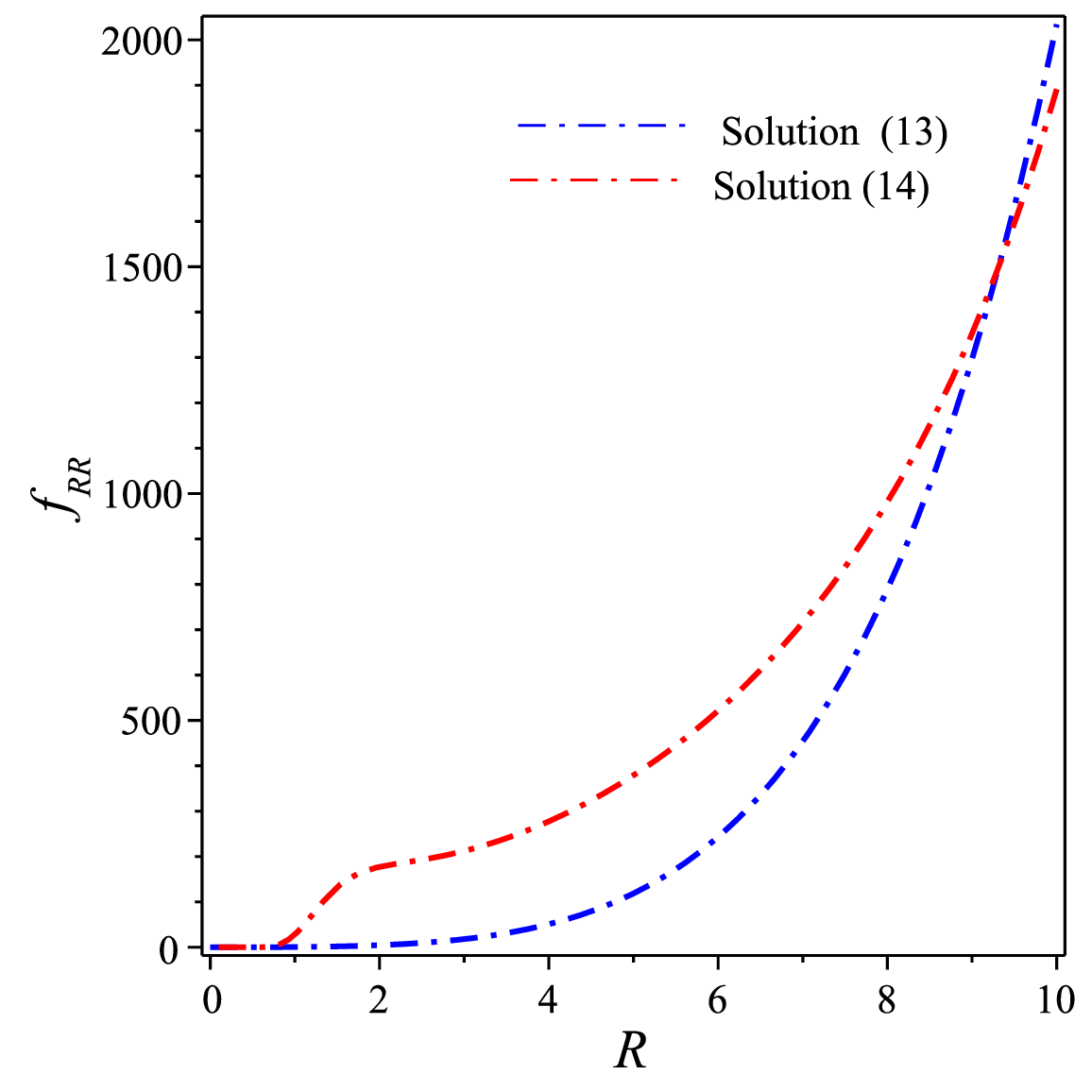}}
\subfigure[~The behavior of the function $f_{RR}$ of the solutions given by Eqs. \eqref{sp111} $\&$ \eqref{sp13d} ]{\label{fig:fRR1}\includegraphics[scale=0.24]{FRMMMM_M_f_RR1.eps}}
\caption[figtopcap]{\small{The plots systematically illustrate: \subref{fig:R} $R$; \subref{fig:fr} and \subref{fig:fr1}  $f(r)$; \subref{fig:fR}  $f_R$; and \subref{fig:fRR} and \subref{fig:fRR1} the second derivative $f_{RR}$. All these results are obtained using the constants $c_0\equiv M=0.3$ and $c_1=c_2=c_3=c_4=c_5=c_6=1$.}}
\label{Fig:1}
\end{figure}
Figure \ref{Fig:1} depicts the characteristics of $R$, $f(r)$, $f_R$, and $f_{RR}$. As can be seen from Figure~\ref{Fig:1}~\subref{fig:R}--\ref{Fig:1}~\subref{fig:fRR}, $R$, $f(r)$, along with the first and second derivatives of $f(R)$, take positive values, thereby fulfilling the Dolgov–Kawasaki stability requirement

{ Before we close this section we should stress that we have obtained new black holes either charged/uncharged that have no well behavior as $r\to 0$ as we discussed in cases {\textbf A},   {\textbf C} and {\textbf D}. As a tentative conclusion of this we can say that these cases suffer singularity as $r \to 0$. These case will become clear when we study their geodesic and shadows which will be done elsewhere.}

%%%%%%%%%%%%%%%%%%%%%%%%%%%%%%%%%%% Section 4 %%%%%%%%%%%%%%%%%%%%%%%%%%%%%%%%%%%%%%%%
\section{Thermodynamics of the BHs}\label{S5}
%%%%%%%%%%%%%%%%%%%%%%%%%%%%%%%%%%%%%%%%%%%%%%%%%%%%%%%%%%%%%%%%%%%%%%%%%%%%%%%%%%%%%%
We are now focused on calculating the thermodynamic quantities of the black hole  solutions derived in the previous section.
%The aim is to verify whether these solutions adhere to the first law of thermodynamics.
This involves determining the black hole's temperature, entropy, and other relevant thermodynamic variables. For the solutions to be physically meaningful, these quantities should satisfy the fundamental relationships prescribed by thermodynamics.
%Specifically, we will check whether changes in the black hole's mass are related to changes in its entropy and temperature, as stipulated by the first law: $dM = T dS$, where \(M\) is the mass, \(T\) is the temperature, and \(S\) is the entropy of the black hole. By computing these thermodynamic quantities and verifying their compliance with the first law, we will assess the thermodynamic consistency of our black hole  solutions within the context of $f(R)$ gravity.
Using the Hawking temperature, which is given by $T=\frac{\kappa }{2\pi }$
(where $\kappa $ is the superficial gravity), we can obtain the Hawking
temperature of  BTZ black hole. For this purpose, by considering $%
k(r)=0$, we can express the mass ($m$) in terms of the radius of the
event horizon ($r_{+}$), the cosmological constant and the charge $q$. Let us analyze this in details for every solution:
%%%%%%%%%%%%%%%%%%%%%%%%%%%%%%%%%%% Section 4 %%%%%%%%%%%%%%%%%%%%%%%%%%%%%%%%%%%%%%%%
\subsection{Thermodynamics of the black hole  (\ref{line1})}\label{S41}
%%%%%%%%%%%%%%%%%%%%%%%%%%%%%%%%%%%%%%%%%%%%%%%%%%%%%%%%%%%%%%%%%%%%%%%%%%%%%%%%%%%%%%
%The asymptotic behavior of the temporal component of the black hole solution (\ref{line1}) is described by Eq. \eqref{line1as}, from which its roots can be derived as:
%\begin{align}\label{hor1}
%&r_{_{\pm}}=\pm\frac{\sqrt {6}}{12}\sqrt {{\frac {\sqrt [3]{12} \left[  {\cal M}^{2/3}+2{c_5}^{2}M \sqrt [3]{12}\Lambda \right] }{\Lambda\sqrt [3]{\cal %M}}}}-\frac{1}{12{\Lambda}{\cal M}^{1/6}}\left\{ \left( 6\sqrt [3 ]{12}\sqrt {{\frac {\sqrt [3]{12} \left( {\cal M} ^{2/3}+2{c_5}^{2}M\sqrt [3]{12}\Lambda \right) }{\Lambda\sqrt [3]{{\cal M}}}}}{\cal M}^{2/3}\right.\right.\nonumber\\
%&\left.\left. \mp{12}^{5/3}\sqrt {{\frac {\sqrt [3]{12} \left(  {\cal M}^{2/3}+2 {c_5}^{2}M\sqrt [3]{12}\Lambda \right) }{\Lambda \sqrt [3]{{\cal M}}}}}{c_5}^{2}M\Lambda+24c_5M\sqrt {6}\sqrt [3]{{\cal M}} \right){\frac {1}{\sqrt {{\sqrt [3]{12} \left(  {\cal M}^{2/3}+2{c_5}^{2}M\sqrt [3]{12}\Lambda \right) }}}}\right\}^{1/2}\,,\end{align}where ${\cal M}=\left( {c_5}^ {2}m \left( m+\sqrt {m \left(m -96{c_5}^{2}\Lambda \right) } \right) \Lambda \right) $ and $r_{_{(\pm)}}$ are the inner and outer radius horizons of the spacetime obtained when $k(r)=0$.
%The metric function for ${\Lambda_1}_{eff}<0$ (AdS spacetime)  has a positive root. For ${\Lambda_1}_{eff}>0$
%(dS spacetime) the metric function is always positive, therefore we will discuss only the AdS case.The horizon is located at
\begin{figure}
\centering
\subfigure[~The behavior of temperature given by Eq. (\ref{temp1})]{\label{fig:temp1}\includegraphics[scale=0.21]{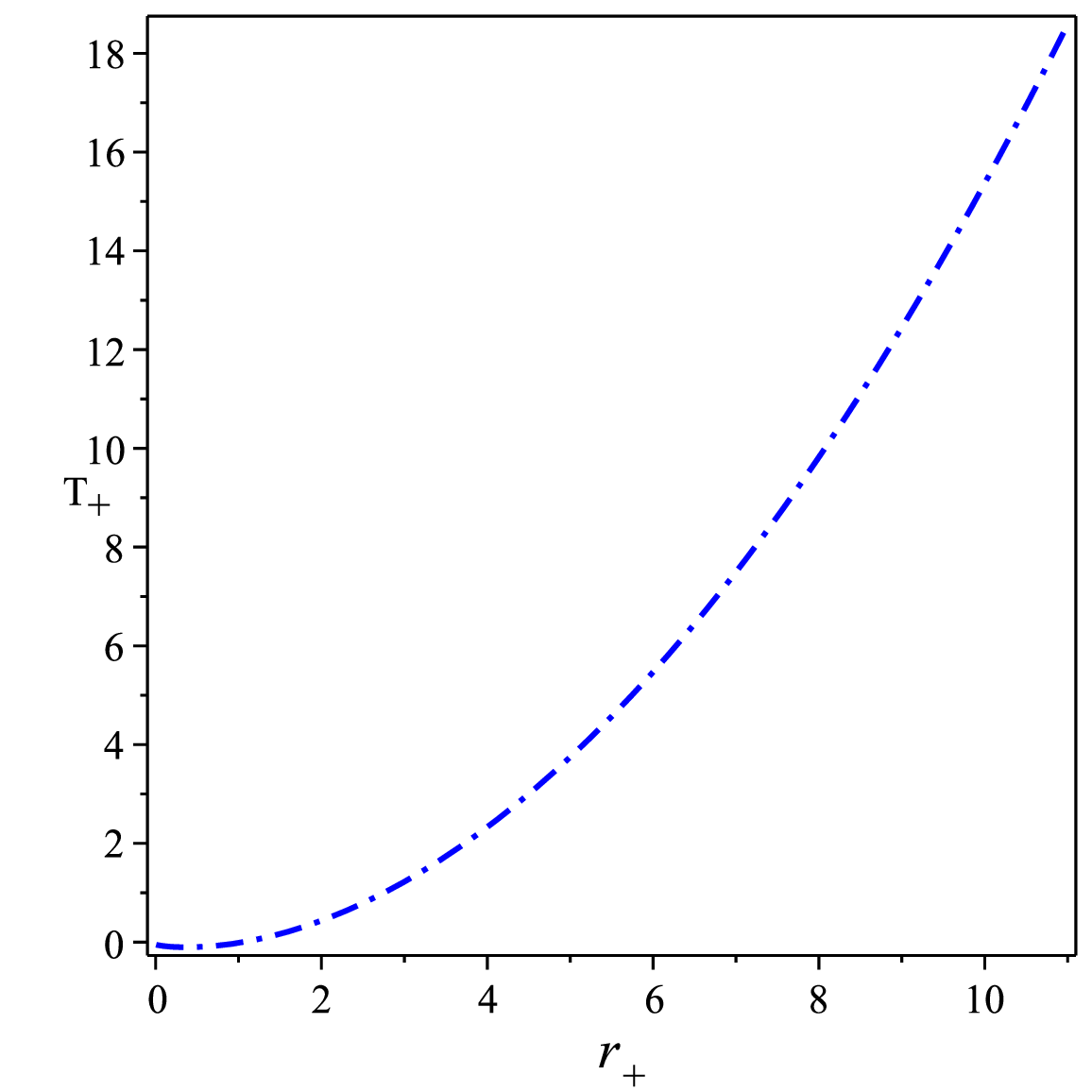}}
\subfigure[~The behavior of temperature given by Eq. (\ref{temp2})]{\label{fig:temp2}\includegraphics[scale=0.21]{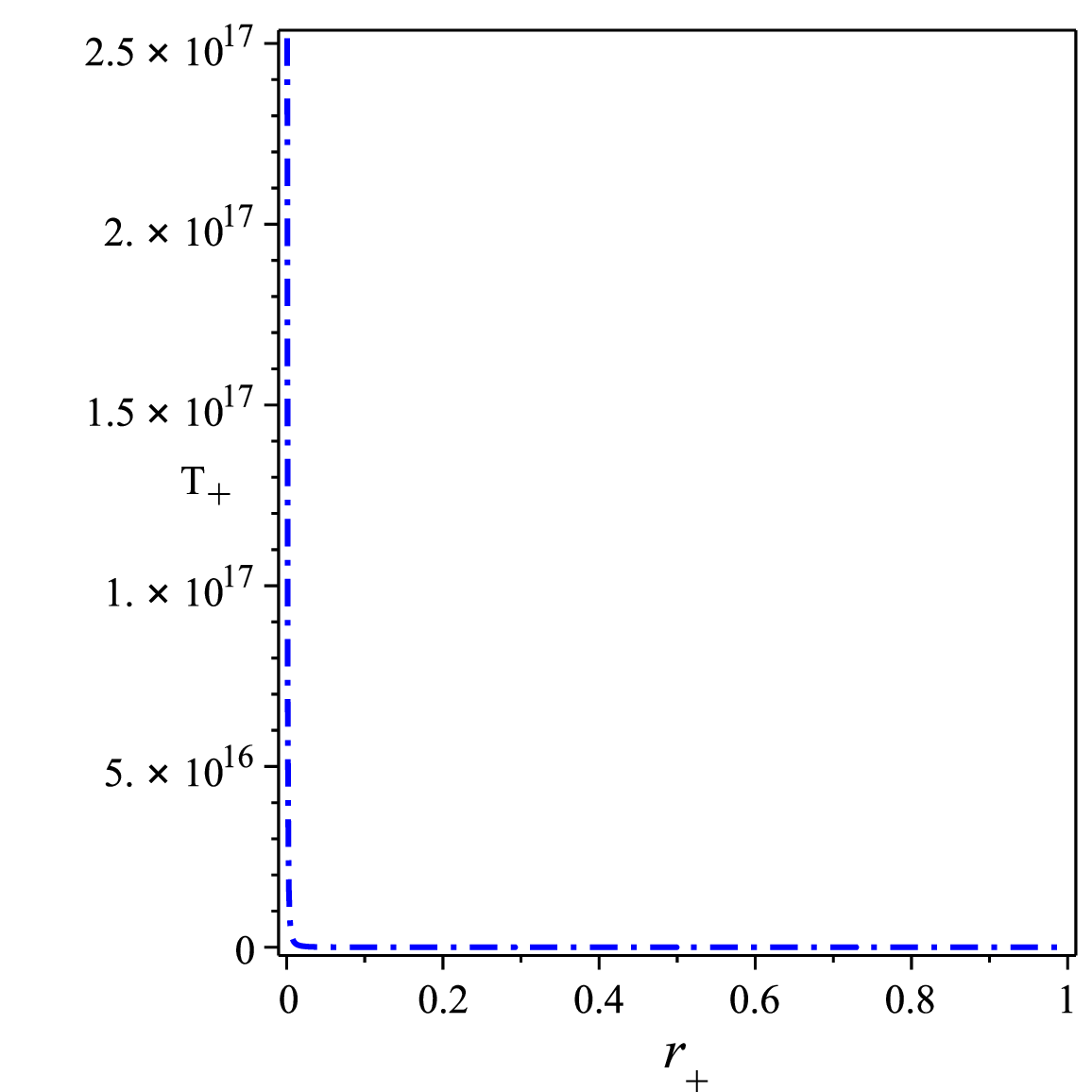}}
\subfigure[~The behavior of temperature given by Eq. (\ref{temp3})]{\label{fig:temp3}\includegraphics[scale=0.21]{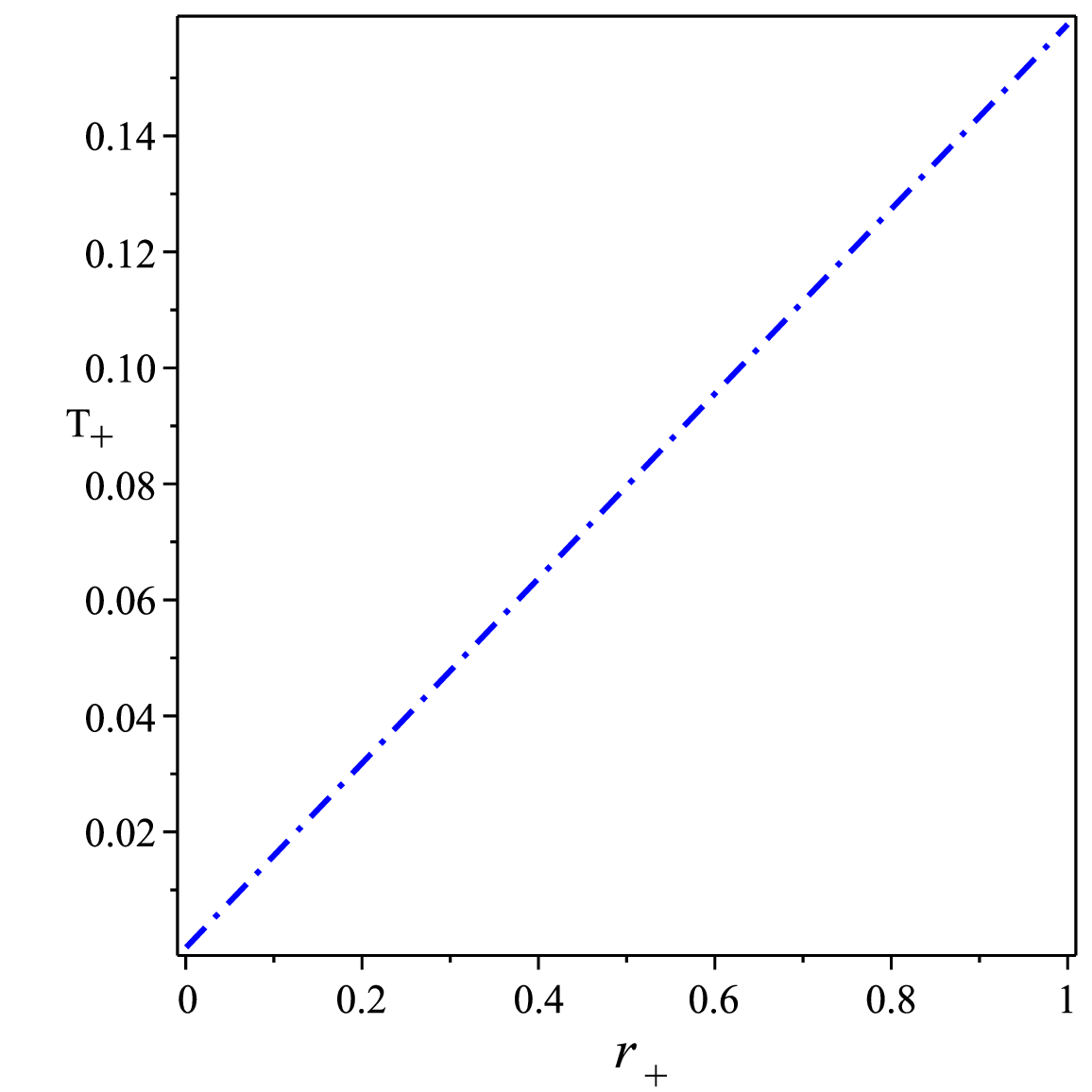}}
\subfigure[~The behavior of temperature given by Eq. (\ref{temp4})]{\label{fig:temp4}\includegraphics[scale=0.21]{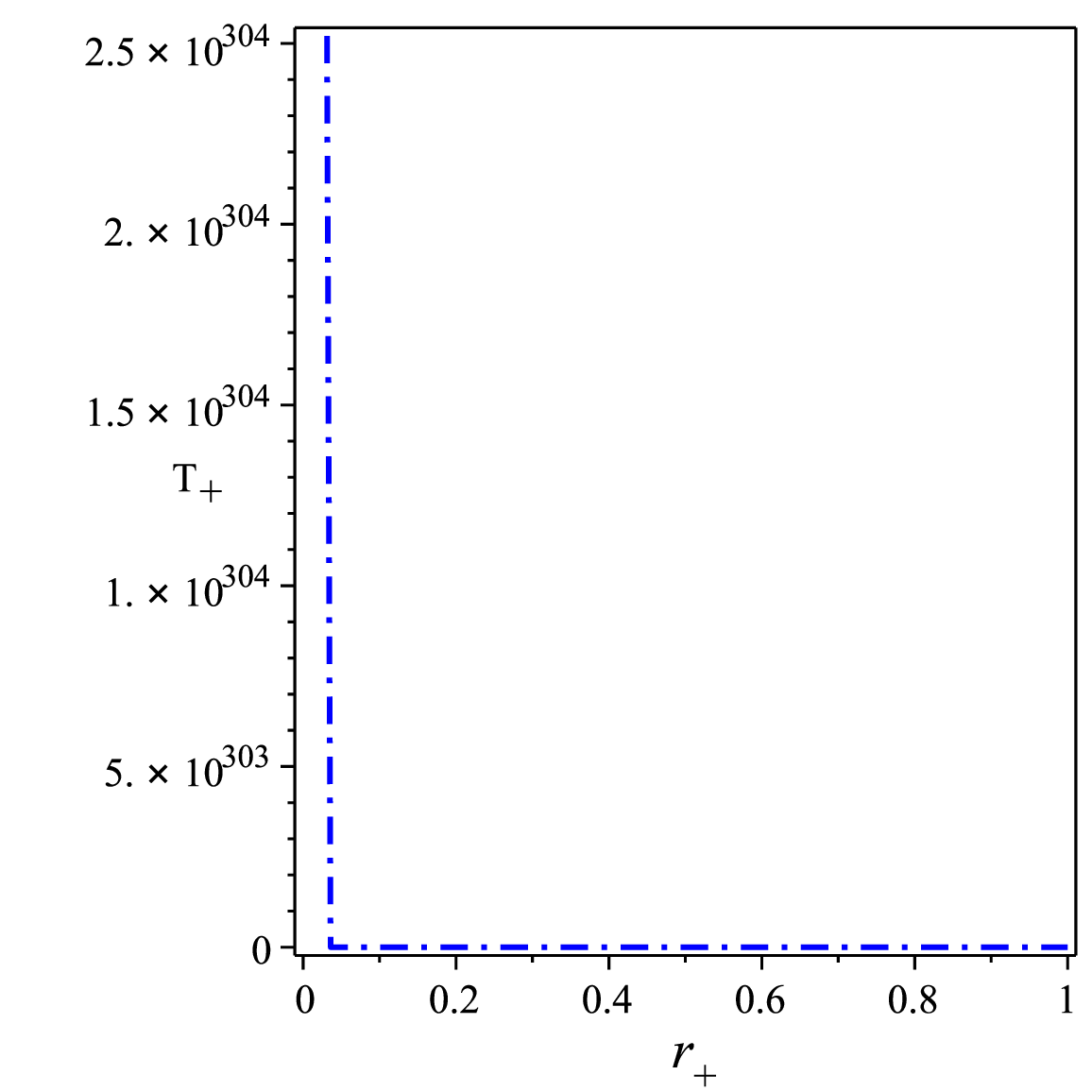}}
\subfigure[~The behavior of entropy given Eq.  (\ref{ent1})]{\label{fig:ent1}\includegraphics[scale=0.21]{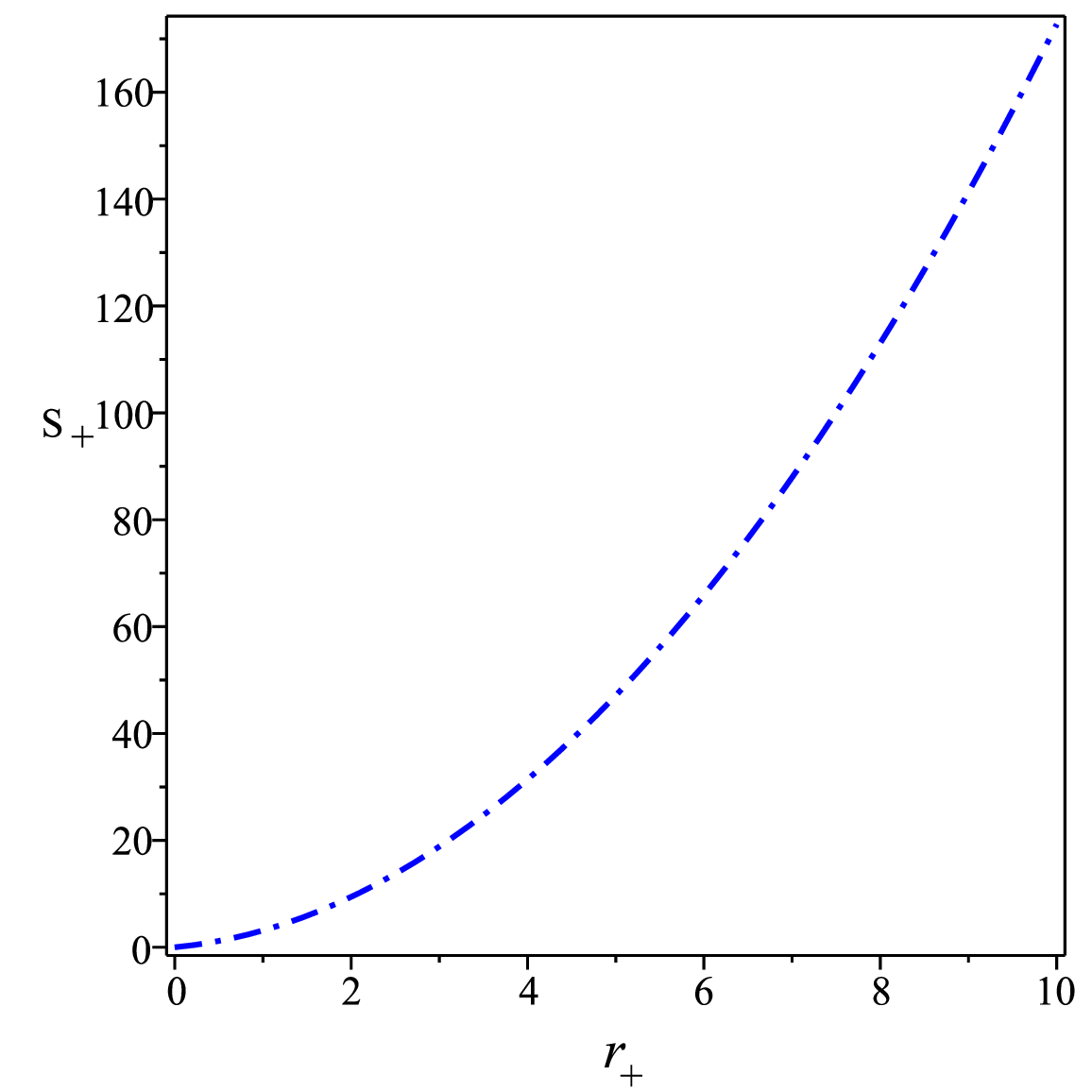}}
\subfigure[~The behavior of entropy given Eq.  (\ref{ent2})]{\label{fig:ent3}\includegraphics[scale=0.21]{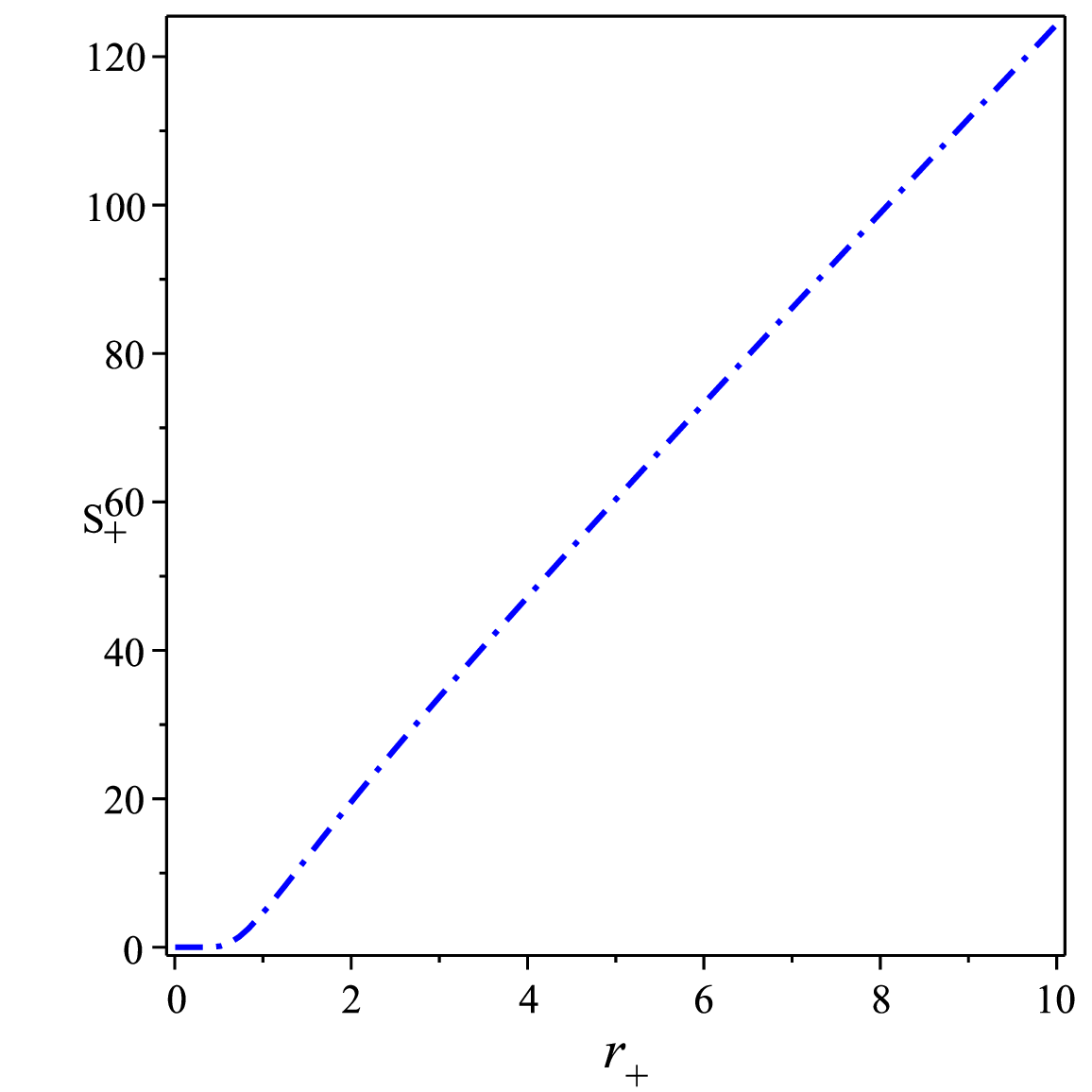}}
\subfigure[~The behavior of entropy given Eq.  (\ref{ent4})]{\label{fig:ent4}\includegraphics[scale=0.21]{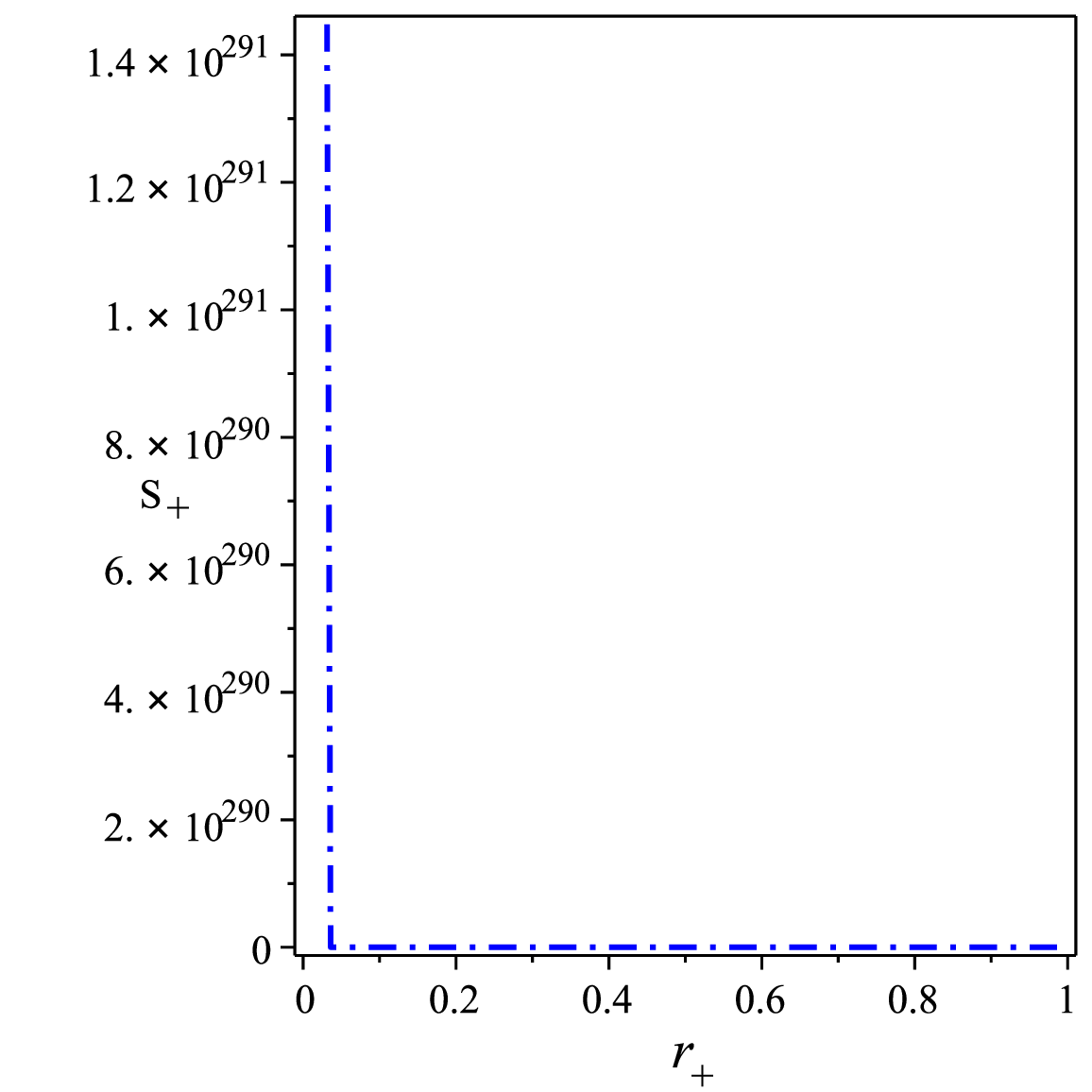}}
\subfigure[~The behavior of the heat capacity  given   (\ref{heat1})]{\label{fig:heat1}\includegraphics[scale=0.21]{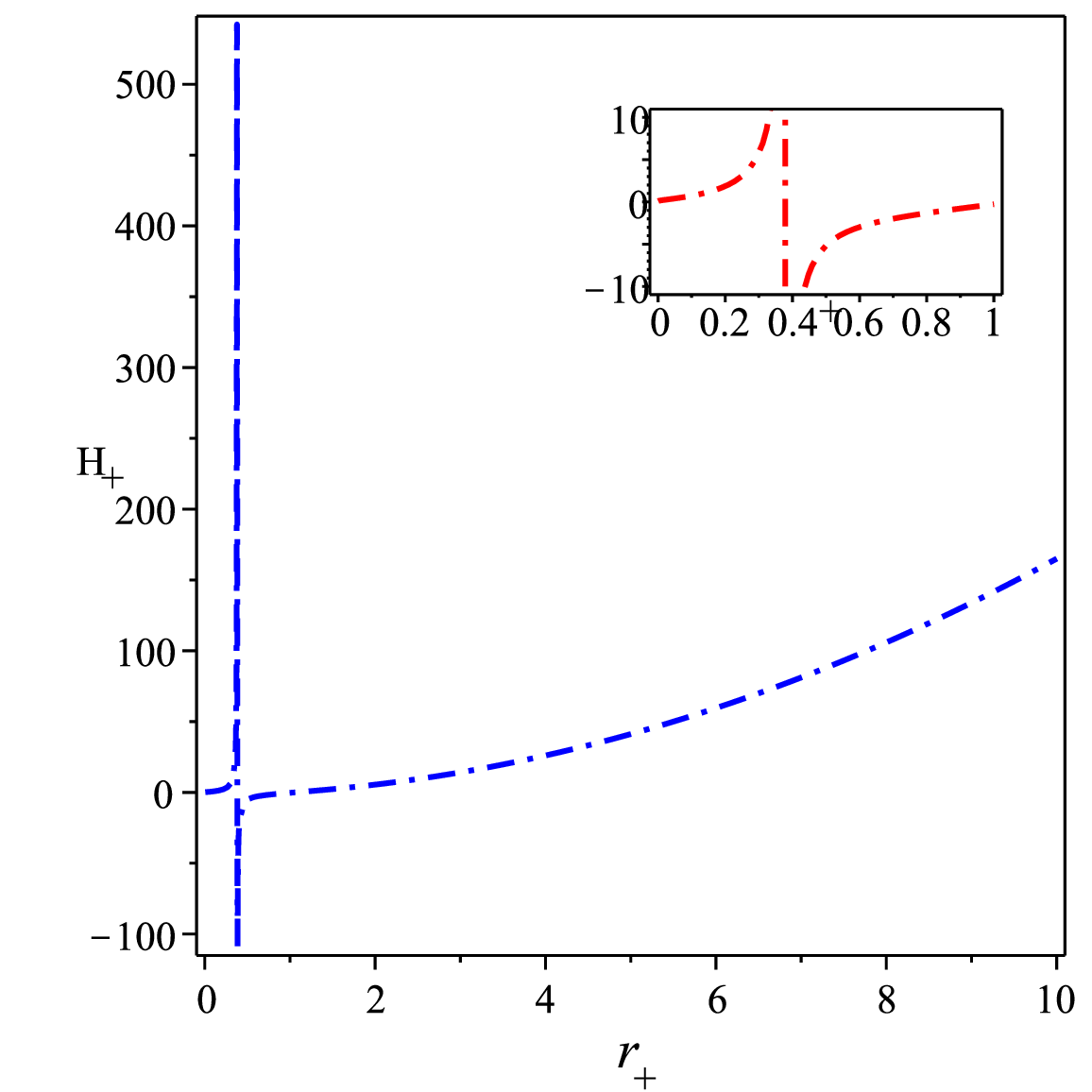}}
\subfigure[~The behavior of the heat capacity  given   (\ref{heat2})]{\label{fig:heat2}\includegraphics[scale=0.21]{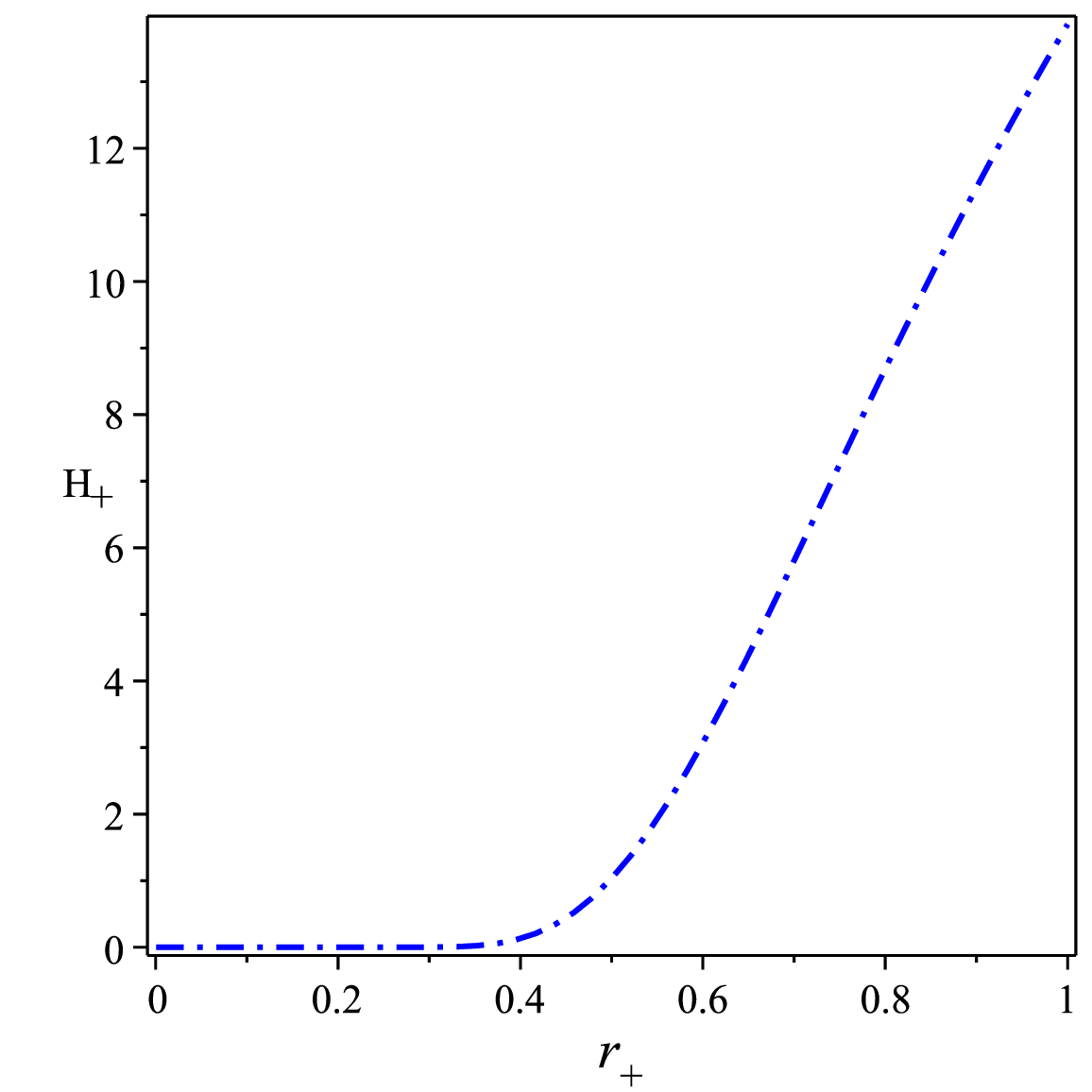}}
\subfigure[~The behavior of the heat capacity  given   (\ref{heat3})]{\label{fig:heat3}\includegraphics[scale=0.21]{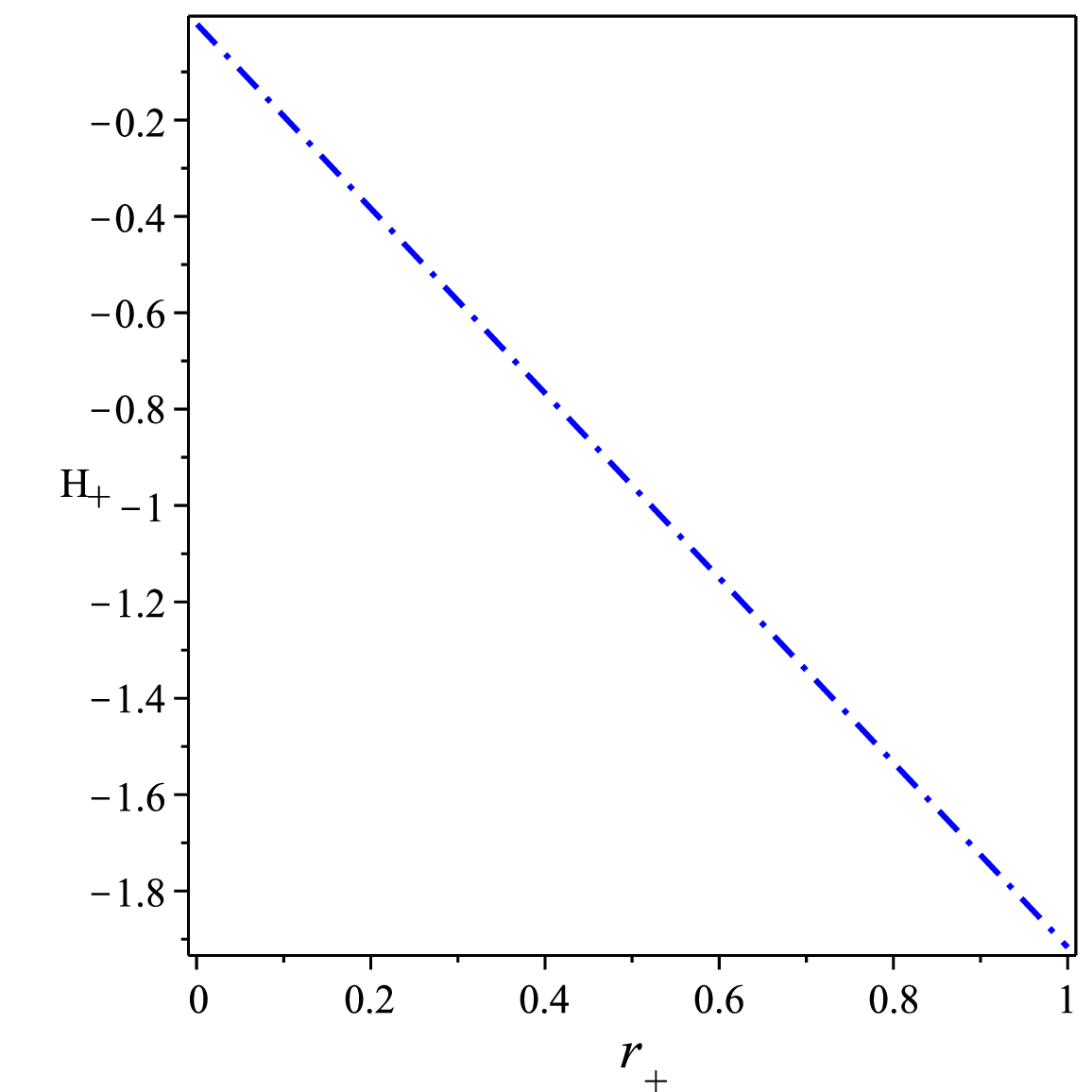}}
\subfigure[~The behavior of the heat capacity  ofEq.~(\ref{heat4})]{\label{fig:heat4}\includegraphics[scale=0.21]{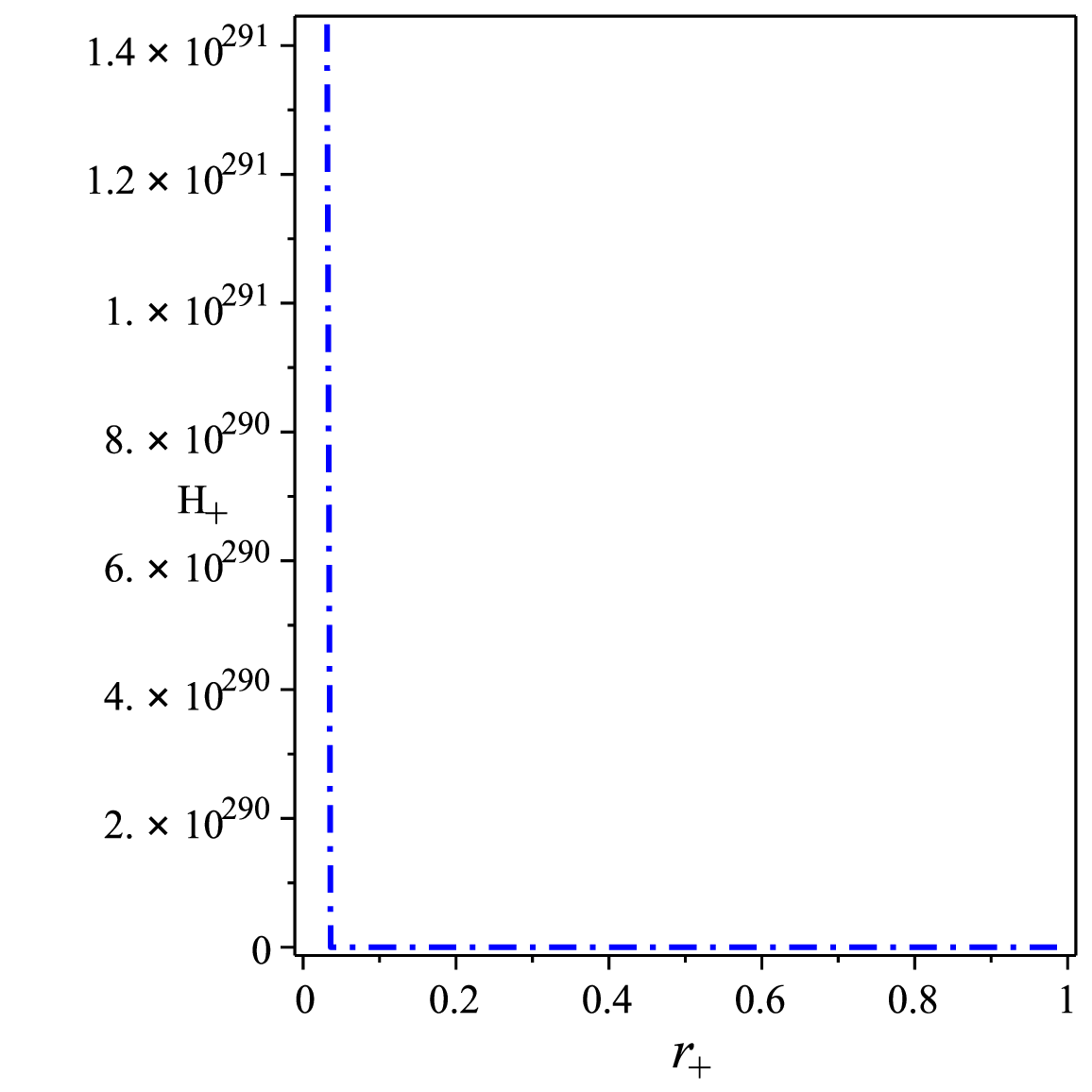}}
\caption[figtopcap]{\small{The  plots include: \subref{fig:temp1}–\subref{fig:temp4}, the Hawking temperature of Eqs.~(\ref{temp1}), (\ref{temp2})–(\ref{temp4}); \subref{fig:ent1}–\subref{fig:ent4}, the entropy from Eqs.~(\ref{ent1}), (\ref{ent2})–(\ref{ent4}); and \subref{fig:heat1}–\subref{fig:heat4}, the heat capacity from Eqs.~(\ref{heat1}), (\ref{heat2})–(\ref{heat4}). These plots are obtained using $M=1$, $c_0=0.01$, $c_1=0.001$, $c_2=-10^{5}$, $c_3=1$, $c_4=1$, and $c_6=-1$. }}
\label{Fig:2}
\end{figure}
We calculate the superficial gravity for the mentioned spacetime (\ref%
{line1as}), which leads to
\begin{equation}
\kappa =\left. \frac{k'(r)}{2}\right\vert _{r=r_{+}},  \label{kGR}
\end{equation}%
where $r_+$ is the radius of the event horizon of the AdS black hole
By considering modified BTZ black hole  given by Eq.~(\ref{line1as}) one can calculate the superficial gravity as:\footnote{Here  we have already used
$r_+$ to replace the parameter $c_0$}.
\begin{equation}
\kappa ={\frac {6\Lambda\,{r}^{4}-2\,m/c_4+3\,m/{c_4}^{2}}{6\pi{r}^{3}}},
\end{equation}%
and so the Hawking temperature leads to \cite{Sheykhi:2012zz,Sheykhi:2010zz,Nashed:2015pga,Hendi:2010gq,Sheykhi:2009pf,Nashed:2014sea,Wang:2018xhw}:
\begin{eqnarray}\label{temp1}
&&T(r_+) = { {\frac {6\Lambda\,{r_+}^{4}-2\,mc_5+3\,m{c_5}^{2}}{12\pi{r_+}^{3}}}}\,\nonumber\\
\end{eqnarray}
As shown in Fig.~\ref{Fig:2} \subref{fig:temp1}, the Hawking temperature derived from Eq.~(\ref{temp1}) is plotted,  which confirms that the temperature does not become negative. The semiclassical Bekenstein–Hawking entropy of the horizons takes the form\footnote{To determine the entropy of BTZ black holes, the area law can be applied as:
\begin{align}
&S=\frac{A}{4}, \quad \mbox{where $A$ is the horizon area and is defined by}\nonumber\\
&A=\left. \int_{0}^{2\pi }\sqrt{g_{\varphi \varphi }}d\varphi \right\vert
_{r=r_{+}}=\left. 2\pi r\right\vert _{r=r_{+}}=2\pi r_{+}, \quad  \mbox{where $g_{\varphi \varphi }=r^{2}$.} \label{AGR}
\end{align}%
 So, the entropy of BTZ black holes is given by
\begin{equation}
S=\frac{\pi r_{+}}{2}.  \label{EntGR}
\end{equation}
}:
\begin{equation}\label{ent1}
S(r_+)=\frac{{\cal A}}{4 } f_R(r_+)=\frac{1}2\pi  r_+f_R(r_+)=\frac{1}2\pi  r_+\left(1+\frac{r_+}{c_5}\right)\,,
\end{equation}
Here ${\cal A}=2\pi r_+$ corresponds to the area of the event horizon. Figure~\ref{Fig:2} \subref{fig:ent1} depicts the entropy, showing that the black hole solution from Eq.~(\ref{ent1}) is always characterized by positive entropy. Lastly, the expression for the heat capacity is   \cite{Zheng:2018fyn,Nashed:2019yto,Kim:2012cma}
\begin{eqnarray} \label{heat1}
&&H(r_+)=T(r_+)\left(\frac{S'(r_+)}{T'(r_+)}\right)=4\,{\frac { \left( 6\Lambda\,{r_+}^{4}-2\,c_5\,Mr_++3\,{c_5}^{2}M \right) \pi\, \left( 2\,r_++c_5\right) r_+
}{c_5\, \left( 6\Lambda\,{r_+}^{4}+4\,c_5\,Mr_+-9\,{c_5}^{2}M \right) }}\,.
\end{eqnarray}
The quantities $S'(r_+)$ and $T'(r_+)$ correspond to the derivatives of entropy and Hawking temperature regarding to the  horizon. The properties of the heat capacity, derived from Eq.~(\ref{heat1}), is shown in Fig.~\ref{Fig:2}. Next, we follow the same strategy for the other solutions, namely Eqs.~\eqref{line3as1}, \eqref{line2as}, and \eqref{line3as}, and we obtain: The Hawking temperatures corresponding to these solutions are given by\footnote{For the second case, the entropy is not plotted since it takes the same form as in the first case.}:
\begin{eqnarray}
&&T(r_+)= {\frac {\Lambda\,{r}^{6}-2Mc_6\,{r}^{2}-8\,M{c_6}^{2}}{2\pi\,{r}^{5}}}\,,\label{temp2}\\
&&T(r_+) \frac{1}{18 {\pi}{r}^{2}} \left[ 9\,\Lambda_1{r}^{3}-3\,mc_5+c_3^{2}
\ln  \left( {\frac {r}{c_5}} \right)  \right]
\,,\label{temp3}\\
&&T(r_+) =-\frac{1}{20{\pi}{r}^{6}{c_6}^{2}} \left[ 6{c_3}^{2}{c_6}^{5}{r}^{2}Ei \left( 1 ,{\frac {c_6}{r}} \right) +10{c_3}^{2}{c_6}^{3}{r}^{4 }Ei \left( 1,{\frac {c_6}{r}} \right) +15{c_3}^{2}{c_6}^{4}{r}^{3}Ei \left( 1,{\frac {c_6}{r}} \right) -10 {r}^{7}\lambda{c_6}^{2}+60M{r}^{2}{c_6}^{5}\right.\nonumber\\
&&\left.+150 M{r}^{3}{c_6}^{4}+100M{r}^{4}{c_6}^{3}+10{e^{-{ \frac {c_6}{r}}}}{c_3}^{2}{c_6}^{7}+52436{e^{-{ \frac {c_6}{r}}}}{c_3}^{2}{r}^{7}+100{e^{-{\frac {c_6}{r}}}}{c_3}^{2}{c_6}^{6}r+2324{e^{-{\frac {c_6}{r}}}}{c_3}^{2}{c_6}^{4}{r}^{3}+540{e^{-{ \frac {c_6}{r}}}}{c_3}^{2}{c_6}^{5}{r}^{2}\right.\nonumber\\
&&\left.+26213{ e^{-{\frac {c_6}{r}}}}{c_3}^{2}{c_6}^{2}{r}^{5}+ 52436{e^{-{\frac {c_6}{r}}}}{c_3}^{2}c_6{r}^{ 6}+8801{e^{-{\frac {c_6}{r}}}}{c_3}^{2}{c_6}^{3 }{r}^{4} \right]\,.\label{temp4}
\end{eqnarray}
\begin{eqnarray}
&&S(r_+)= 4\, \pi^2{r} {e^{-{\frac {c_6}{{
r^2}}}}}\,,\label{ent2}\\
&&S(r_+) =4\, \pi^2{r} {e^{-{\frac {c_6}{{
r}}}}}\, \,.\label{ent4}
\end{eqnarray}

\begin{eqnarray}
&&H(r_+)={\frac { 4\left( \Lambda\,{r}^{6}-2\,c_6\,{r}^{2}-8\,M{c_6}^{2} \right) \pi\, \left( 2\,r+c_5 \right) r}{c_5\,
 \left( \Lambda\,{r}^{6}+6c_6\,{r}^{2}+40\,M{c_6}^{2}
 \right) }}\,,\label{heat3}\\
&&H(r_+)=\frac{4 \left( {r}^{2}+2c_6 \right) {e^{-{\frac {c_6}{{ r}^{2}}}}}\pi \left(9\Lambda_1{r}^{3}-3M{r}^{2}c_5-c_3^2\ln  \left( {\frac {r}{c_5}} \right)\right)}{ {r} \left[9 \Lambda_1{r}^{3}-c_3^2+6Mc_5+2c_3^2\ln  \left( {\frac {r}{c_5}} \right) \right] }\label{heat2}\,,\\
&&H(r_+)=-4\pi \left[ 6{c_3}^{2}{c_6}^{5}{r}^{2}Ei \left( 1,{ \frac {c_6}{r}} \right) +10{c_3}^{2}{c_6}^{3}{r}^{4}Ei \left( 1,{\frac {c_6}{r}} \right) +15{c_3}^{2}{c_6}^{4}{r}^{3}Ei \left( 1,{\frac {c_6}{r}} \right) -10 {r}^{7}\Lambda{c_6}^{2}+60M{r}^{2}{c_6}^{5}+150 M{r}^{3}{c_6}^{4}\right.\nonumber\\
 &&\left.+100M{r}^{4}{c_6}^{3}+{e^{-{ \frac {c_6}{r}}}}\left\{10{c_3}^{2}{c_6}^{7}+52436{c_3}^{2}{r}^{7}+100{c_3}^{2}{c_6}^{6}r+2324{c_3}^{2}{c_6}^{4}{r}^{3}+540{c_3}^{2}{c_6}^{5}{r}^{2}
 +26213{c_3}^{2}{c_6}^{2}{r}^{5}+ 52436{c_3}^{2}c_6{r}^{ 6}\right.\right.\nonumber\\
 &&\left.\left.+8801{c_3}^{2}{c_6}^{3 }{r}^{4}\right\}\right] {e^{-{\frac {c_6}{r}}}} \left( r+c_6 \right] r \left[ -52436{e^{-{\frac {c_6}{r}}}}{ c_3}^{2}{r}^{8}+240M{r}^{3}{c_6}^{5}+450M{r}^{4}{c_6 }^{4}+200M{r}^{5}{c_6}^{3}+10{r}^{8}\Lambda{c_6} ^{2}-40{e^{-{\frac {c_6}{r}}}}{c_3}^{2}{c_6}^{7 }r\right.\nonumber\\
 &&\left.-10{c_6}^{8}{e^{-{\frac {c_6}{r}}}}{c_3}^{2}+ 24{c_3}^{2}{c_6}^{5}{r}^{3}Ei \left( 1,{\frac {c_6}{r}} \right) -170{c_3}^{2}{c_6}^{5}{r}^{3}{e^{-{ \frac {c_6}{r}}}}+20{c_3}^{2}{c_6}^{3}{r}^{5}Ei \left( 1,{\frac {c_6}{r}} \right) -8621{c_3}^{2}{c_6 }^{3}{r}^{5}{e^{-{\frac {c_6}{r}}}}\right.\nonumber\\
 &&\left.+45{c_3}^{2}{c_6}^{4}{r}^{4}Ei \left( 1,{\frac {c_6}{r}} \right) - 1844{c_3}^{2}{c_6}^{4}{r}^{4}{e^{-{\frac {c_6}{ r}}}}-52436c_6{r}^{7}{e^{-{\frac {c_6}{r}}}}{c_3}^{2}-26223{e^{-{\frac {c_6}{r}}}}{c_3}^{2}{c_6}^{2}{r}^{6}-40{e^{-{\frac {c_6}{r}}}}{c_3}^{2}{c_6}^{6}{r}^{2} \right]^{-1}\,.\label{heat4}
\end{eqnarray}

%%%%%%%%%%%%%%%%%%%%%%%%%%%%%%%%%%%%%%%%%%%%%%%%%%%%%%%%%%%%%%%%%%%%%%%%%%%%%%%%%%%%%%
\section{ Discussion and conclusions }\label{S77}
Through our analysis of $(2+1)$-dimensional $f(R)$ gravity, we have identified a variety of black hole geometries along with their associated thermodynamic characteristics.  Through careful mathematical exploration, we have derived novel BH solutions characterized by their unique dependence on the $f(R)$ function, which itself has been shown to behave as polynomial functions. These solutions exhibit remarkable behaviors, including the ability to display asymptotically Anti-de Sitter (AdS) or de Sitter (dS) characteristics, despite the absence of an explicit cosmological constant in the  equations of motion of $f(R)$ gravity. Moreover, the Ricci scalar of these solutions is not constant which ensure that those black holes are new in the frame of $(2+1)$-dimensional geometry.

Notably, the study has highlighted the importance of the $f(R)$ function's form and its parameters in determining the black holes properties, including its thermodynamic stability. To show the stability of these black holes we derive the related form of $f(R)$, $f_R$ and $f_{RR}$ of each solution showing that all of these quantities have a positive pattern thereby confirming that the Dolgov–Kawasaki stability condition is obeyed  \cite{DeFelice:2010aj,Bertolami:2009cd,Faraoni:2006sy,Cognola:2007zu}.  We also calculate the heat capacity of each solution and demonstrated that all heat capacities of these black holes have positive pattern which mean that these solution are stable \cite{Hendi:2014mba}.

Overall, this work has not only expanded the existing body of knowledge on black hole solutions in alternative theories of gravity but has also provided valuable insights into the fundamental properties of BHs in (2+1)dimensions. Future research may explore more complex scenarios, including rotating BHs or those with additional matter fields, to further probe the richness of $f(R)$ gravity and its implications for black hole physics. Our findings underscore the potential of $f(R)$ gravity to offer new perspectives on the nature of black holes and the fabric of spacetime itself, inviting further exploration and theoretical development.

{ Since Hawking's discovery of black hole thermal radiation \cite{Hawking:1974rv,Hawking:1975vcx,Hawking:1975vcx}, the definition of black hole entropy within a generally covariant theory has become well established. In particular, for Einstein gravity minimally coupled to matter, the entropy is given by one quarter of the horizon area. This area law has been extended to the Wald entropy formula to account for more complex couplings or higher-order curvature terms. When applied to static black holes with spherical, toroidal, or hyperbolic symmetries, it was found that Horndeski terms do not contribute to the Wald entropy. This might suggest that the entropy remains one quarter of the horizon area. However, this is not the case. A detailed examination of the Wald procedure \cite{Wald:1993nt,Iyer:1994ys} reveals that in Horndeski gravity, there is an additional contribution to the entropy not captured by the standard Wald formula \cite{Feng:2015oea}. Whether this result applies to the black holes derived in the present study remains an open question and will be investigated in future work.}
\section{acknowledgment}
I am  greatly indebted to the anonymous referee for the
constructive comments to improve the presentation of this
work.
%%%%%%%%%%%%%%%%%%%%%%%%%%%%%%%%%%%%%%%%%%%%%%%%%%%%%%%%%%%%%%%%%%%%%%%%%%%%%%%%%%%%%%
%\bibliographystyle{apsrev}
%\bibliography{JRPHSRef}

\end{document}